\documentclass[11pt,draft,onecolumn]{IEEEtran}

\usepackage{graphicx}
\usepackage{epsfig}
\usepackage{amsmath}
\usepackage{amssymb}
\usepackage{subfigure}
\usepackage{wrapfig}
\usepackage{times,color}
\usepackage{cite,footnote,xspace,syntonly,algorithm,algorithmic,bm}
\input{mysymbol.sty}
\newcommand{\calG}{{\cal G}}
\newcommand{\calR}{{\cal R}}
\newcommand{\calE}{{\cal E}}
\newcommand{\calN}{{\cal N}}

\newcommand{\calbbv}{\boldsymbol{{\mathit{v}}}}

\newcommand{\rpr}{r^\prime}

\long\def\symbolfootnote[#1]#2{\begingroup
\def\thefootnote{\fnsymbol{footnote}}
\footnote[#1]{#2}\endgroup} \psfull

\begin{document}
\title{\huge Group-Lasso on  Splines for Spectrum Cartography$^\dag$}

\author{{\it Juan~A.~Bazerque, Gonzalo~Mateos
         and Georgios~B.~Giannakis~(contact author)$^\ast$}}

\markboth{IEEE TRANSACTIONS ON SIGNAL PROCESSING (SUBMITTED)}
\maketitle \maketitle \symbolfootnote[0]{$\dag$ Work in this paper
was supported by the NSF grants CCF-0830480 and ECCS-0824007. Part
of the paper appeared in the {\it Proc. of the 43rd Asilomar
Conference on Signals, Systems, and Computers}, Pacific Grove, CA,
Nov. 1-4, 2009.} \symbolfootnote[0]{$\ast$ The authors are with
the Dept. of Electrical and Computer Engineering, University of
Minnesota, 200 Union Street SE, Minneapolis, MN 55455. Tel/fax:
(612)626-7781/625-4583; Emails:
\texttt{\{bazer002,mate0058,georgios\}@umn.edu}}

\vspace*{-80pt}
\begin{center}
\small{\bf Submitted: }\today
\end{center}
\vspace*{10pt}

\thispagestyle{empty}\addtocounter{page}{-1}
\begin{abstract}
The unceasing demand for continuous situational awareness calls
for innovative and large-scale signal processing algorithms,
complemented by collaborative and adaptive sensing platforms to
accomplish the objectives of layered sensing and control. Towards
this goal, the present paper develops a spline-based approach to
field estimation, which relies on a basis expansion model of the
field of interest. The model entails known bases, weighted by
generic functions estimated from the field's noisy samples. A
novel field estimator is developed based on a regularized
variational least-squares (LS) criterion that yields
finitely-parameterized (function) estimates spanned by thin-plate
splines. Robustness considerations motivate well the adoption of
an overcomplete set of (possibly overlapping)  basis functions,
while a sparsifying regularizer augmenting the LS cost endows the
estimator with the ability to select a few of these bases that
``better'' explain the data. This parsimonious field
representation becomes possible, because the sparsity-aware
spline-based method of this paper induces a group-Lasso estimator
for the coefficients of the thin-plate spline expansions per
basis. A distributed algorithm is also developed to obtain the
group-Lasso estimator using a network of wireless sensors, or,
using multiple processors to balance the load of a single
computational unit. The novel spline-based approach is motivated
by a \textit{spectrum cartography} application, in which a set of
sensing cognitive radios collaborate to estimate the distribution
of RF power in space and frequency.  Simulated tests corroborate
that the estimated power spectrum density atlas yields the desired
RF state awareness, since the maps reveal spatial locations where
idle frequency bands can be reused for transmission, even when
fading and shadowing effects are pronounced.
\end{abstract}

\begin{keywords}
Sparsity, splines, (group-)Lasso, field estimation, cognitive
radio sensing, optimization.
\end{keywords}
%

\newpage

\section{Introduction}\label{sec:introduction}
Well-appreciated as a tool for  field estimation, thin-plate
(smoothing) splines find application in  areas as diverse as
climatology \cite{wahba_weather}, image processing
\cite{Chui00anew}, and neurophysiology \cite{Perrin87}. Spline-based
field estimation involves approximating a deterministic map
$g:\mathbb R^n\to \mathbb R$ from a finite number of its noisy data
samples, by minimizing a variational least-squares (LS) criterion
regularized with a smoothness-controlling functional. In the dilemma
of trusting a model versus trusting the data, splines favor the
latter since only a mild regularity condition is imposed on the
derivatives of $g$, which is otherwise treated as a generic
function. While this generality is inherent to the variational
formulation, the smoothness penalty renders the estimated map unique
and finitely parameterized~\cite[p. 85]{duchon},~\cite[p.
31]{wahba}. With the variational problem solution expressible by
polynomials and specific kernels, the aforementioned map
approximation task reduces to a parameter estimation problem.
Consequently, thin-plate splines operate as a reproducing kernel
Hilbert space (RKHS) learning machine in a suitably defined
(Sobolev) space~\cite[p. 34]{wahba}.

Although splines emerge as variational LS estimators of
\emph{deterministic} fields, they are also connected to classes
of estimators  for \emph{random} fields. The first class
assumes that estimators are linearly related to the measured
samples, while the second one assumes that fields are Gaussian
distributed. The first corresponds to the Kriging method while
the second to the Gaussian process model; but in both cases one
deals with a best linear unbiased estimator
(BLUE)~\cite{Stein_Kriging}. Typically, wide sense stationarity
is assumed for the field's spatial correlation needed to form
the BLUE. The so-termed generalized covariance model adds a
parametric nonstationary term comprising known functions
specified a priori~\cite{matheron}. Inspection of the BLUE
reveals that if the nonstationary part is selected to comprise
polynomials, and the spatial correlation is chosen to be the
splines kernel, then the Kriging, Gaussian process, and
spline-based estimators coincide~\cite[p. 35]{wahba}.

Bearing in mind this unifying treatment of deterministic and
random fields, the main subjects of this paper are spline-based
estimation, and the practically motivated \emph{sparse} (and
thus parsimonious) description of the wanted field. Toward
these goals, the following basis expansion model (BEM)  is
adopted for the target map
\begin{equation}\label{eq:basis-expansion2}
\Phi(\mathbf x, f)=\sum_{\nu=1}^{N_b}g_\nu(\mathbf x)b_\nu(f)
\end{equation}
with $\mathbf x\in\mathbb R^2$, $f\in \mathbb R$, and the
$L_2-$norms $\{||b_\nu(f)||_{L_2}=1\}_{\nu=1}^{N_b}$ normalized to
unity.

The bases $\{b_\nu(f)\}_{\nu=1}^{N_b}$ are preselected, and the
functions $g_\nu(\bf x)$ are to be estimated based on noisy
samples of $\Phi$.
This way, the model-versus-data balance is calibrated by
introducing a priori knowledge on the dependence of the map $\Phi$
with respect to (w.r.t.) variable $f$, or more generally a group
of variables, while trusting the data to dictate the functions
$g_\nu(\bf x)$ of the remaining variables $\mathbf x$.

Consider selecting  $N_b$ basis functions using the \emph{basis
pursuit} approach~\cite{basis_pursuit}, which entails an extensive
set of bases thus rendering $N_b$ overly large and the model
overcomplete.
This motivates augmenting the variational LS problem with a
suitable sparsity-encouraging penalty, which endows the map
estimator with the ability to discard factors $g_\nu(\mathbf x)
b_\nu(f)$ in \eqref{eq:basis-expansion2}, only keeping a few bases
that ``better'' explain the data.
This attribute is inherited  because the novel sparsity-aware
spline-based method of this paper induces a group-Lasso estimator
for the coefficients of the optimal finitely-parameterized $
g_\nu$. Group-Lasso estimators are known to set groups of weak
coefficients to zero (here the $N_b$ groups  associated with
coefficients per $g_\nu$), and outperform the sparsity-agnostic LS
estimator by capitalizing on the sparsity
present~\cite{yuan_group_lasso},~\cite{puig_hero}. An iterative
 group-Lasso algorithm is developed that yields closed-form estimates per
 iteration.
 A distributed version of this algorithm is also
introduced for  data  collected by cooperating sensors, or, for
computational load-balancing of multiprocessor architectures. A
related approach to model selection in nonparametric regression is
the component selection and smoothing operator
(COSSO)~\cite{cosso}. Different from the approach followed here,
 COSSO is limited to smoothing-spline, analysis-of-variance
models, where the target function is assumed to be expressible by a
superposition of \textit{orthogonal} component functions. Compared
to the single group-Lasso estimate here,  COSSO entails an iterative
algorithm, which alternates through a sequence of smoothing
spline\cite[p. 151]{elements_of_statistics} and nonnegative
garrote~\cite{Breiman_garrote_1995} subproblems.

The motivation behind the BEM in \eqref{eq:basis-expansion2} comes
from our interest in spectrum cartography for wireless
\emph{cognitive radio} (CR) networks, a \emph{sensing} application
that  serves as an illustrating paradigm throughout the paper.  CR
technology holds great promise to address fruitfully the perceived
dilemma of bandwidth under-utilization versus spectrum scarcity,
which has rendered fixed-access communication networks
inefficient.
 Sensing the ambient interference spectrum is of paramount importance to the operation of CR
networks, since it enables spatial frequency reuse and allows for
dynamic spectrum allocation; see,
e.g.,~\cite{ganesan08},~\cite{mishra06} and references therein.
Collaboration among CRs can markedly improve the sensing
performance~\cite{collaborative_tutorial}, and is key to revealing
opportunities for spatial frequency reuse~\cite{popovski}.
Pertinent existing approaches have mostly relied on detecting
spectrum occupancy per radio, and do not account for
spatio-temporal changes in the radio frequency (RF) ambiance,
especially at intended receiver(s) which may reside several hops
away from the sensed area.

The impact of this paper's novel field estimators to CR networks
is a collaborative sensing scheme whereby receiving CRs cooperate
to estimate the distribution of power in space $\mathbf{x}$ and
frequency $f$, namely the power spectrum density (PSD) map
$\Phi(\mathbf{x},f)$ in \eqref{eq:basis-expansion2}, from local
periodogram measurements. The estimator need not be extremely
accurate, but precise enough to identify spectrum holes. This
justifies adopting the known bases to capture the PSD frequency
dependence in \eqref{eq:basis-expansion2}. As far as the spatial
dependence is concerned, the model must account for path loss,
fading, mobility, and shadowing effects, all of which vary with
the propagation medium. For this reason, it is prudent to let the
data dictate the spatial component of \eqref{eq:basis-expansion2}.
Knowing the spectrum at any location allows remote CRs to reuse
dynamically idle bands. It also enables CRs to adapt their
transmit-power so as to minimally interfere with licensed
transmitters.
The spline-based PSD map here provides an alternative
to~\cite{bazerque}, where known bases  are used both in space and
frequency. Different from~\cite{cartography} and \cite{bazerque},
the field estimator here does not presume a spatial covariance
model or pathloss channel model. Moreover, it captures general
propagation characteristics including both shadowing and fading;
see also~\cite{Kim09}.


\noindent\textit{Notation:} Bold uppercase letters will denote
matrices, whereas bold lowercase letters will stand for column
vectors. Operators $\otimes$, $(.)^{\prime}$, $\mbox{tr}(.)$,
$\mbox{rank}(.)$, $\mbox{bdiag}(.)$, $E[\cdot]$ will denote
Kronecker product, transposition, matrix trace, rank, block
diagonal matrix and expectation, respectively; $|.|$ will be used
for the cardinality of a set, and the magnitude of a scalar. The
$L_2$ norm of  function $b:\mathbb R \to \mathbb R$ is
$||b||_{L_2}^2:=\int_{-\infty}^\infty b^2(f) df$, while the
$\ell_p$ norm of vector $\mathbf{x}\in\mathbb{R}^{p}$ is
$\|\mathbf{x}\|_p:=\left(\sum_{i=1}^p|x_i|^p\right)^{1/p}$ for
$p\geq 1$; and
$\|\mathbf{M}\|_F:=\sqrt{\mbox{tr}\left(\mathbf{M}\mathbf{M}^{\prime}\right)}$
is the matrix Frobenious norm. Positive definite matrices will be
denoted by $\bbM\succ\mathbf{0}$. The $p\times p$ identity matrix
will be represented by $\mathbf{I}_{p}$, while $\mathbf{0}_{p}$
will denote the $p\times 1$ vector of all zeros, and
$\mathbf{0}_{p\times q}:=\mathbf{0}_{p}\mathbf{0}_{q}^{\prime}$.
The $i$-th vector in the canonical basis for $\mathbb{R}^p$ will
be denoted by $\mathbf{e}_{p,i}$, $i=1,\ldots,p$.

%
%

\section{BEM for Spectrum Cartography}\label{sec:modelling}
\begin{figure}[h]
\begin{center}
\includegraphics[width=8cm]{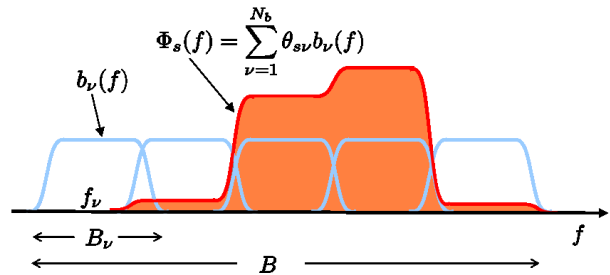}
\vspace{-0.1cm} \caption{Expansion with overlapping raised cosine
pulses.}\label{fig:frequency_basis}
\end{center}
\end{figure}

Consider a set of $N_s$ sources transmitting signals
$\{u_s(t)\}_{s=1}^{N_s}$ using portions of the overall bandwidth
$B$. The objective of revealing which of these portions
(sub-bands) are available for new systems to transmit, suggests
that the PSD estimate sought does not need to be super accurate.
This motivates modeling the transmit-PSD of each $u_s(t)$ as
\begin{equation}\label{eq:intro-frequency-basis-expansion-psd_i}
\Phi_s(f)=\sum_{\nu =1}^{N_b}\theta_{s\nu } b_{\nu }(f),~~
s=1,\ldots,N_s
\end{equation}
where the basis $b_\nu(f)$ is centered at frequency $f_\nu,\
\nu=1,\ldots,N_b$.
The example depicted in Fig. \ref{fig:frequency_basis} involves
(generally \emph{overlapping}) raised cosine bases with support
$B_\nu=[f_\nu-(1+\rho)/2T_s,f_\nu+(1+\rho)/2T_s]$, where $T_s$ is
the symbol period, and $\rho$ stands for the roll-off factor. Such
bases can model transmit-spectra of e.g., multicarrier systems. In
other situations, power spectral masks may dictate sharp
transitions between contiguous sub-bands, cases in which
non-overlapping rectangular bases may be more appropriate. All in
all, the set of bases should be selected to accommodate a priori
knowledge about the PSD.


The power transmitted by source $s$ will propagate to the location
$\mathbf{x}\in\mathbb{R}^2$ according to a generally unknown
spatial loss function $l_s(\mathbf{x}): \mathbb{R}^2\to
\mathbb{R}$. The propagation model  $l_s(\mathbf{x})$ not only
captures frequency-flat deterministic pathloss, but also
stationary, block-fading and even frequency-selective Rayleigh
channel effects, since their statistical moments do not depend on
the frequency variable. In this case, the following vanishing
memory assumption is required on the transmitted signals for the
spatial receive-PSD $ \Phi(\mathbf x,f)$   to be factorizable as
$l_s(\mathbf{x})\Phi_s(f)$; see~\cite{bazerque} for further
details.

\noindent\textbf{(as)} \textit{Sources $\{u_s(t)\}_{s=1}^{N_s}$
are stationary, mutually uncorrelated, independent of the
channels, and have vanishing correlation per coherence interval;
i.e., $r_{ss}(\tau):=E[u_s(t+\tau)u_s(t)]=0,$
$\forall\:|\tau|>T_c-L$, where $T_c$ and $L$ represent the
coherence interval and delay spread of the channels,
respectively.}

Under (as), the contribution of source $s$ to the PSD at point
$\mathbf{x}$ is $l_s(\mathbf{x})\sum_{\nu =1}^{N_b}\theta_{s\nu}
b_{\nu}(f)$; and the PSD due to all sources received at
$\mathbf{x}$ will be given by
$\Phi(\mathbf{x},f)=\sum_{s=1}^{N_s}l_s(\mathbf{x})\sum_{\nu
=1}^{N_b}\theta_{s\nu} b_{\nu }(f)$. Such a model can be
simplified by defining the function
$g_\nu(\mathbf{x}):=\sum_{s=1}^{N_s}\theta_{s\nu}
l_s(\mathbf{x})$. With this definition and upon exchanging the
order of summation, the spatial PSD model takes the form in
\eqref{eq:basis-expansion2},
%
%
where functions $\{g_\nu(\mathbf{x})\}_{\nu=1}^{N_b}$ are to be
estimated. They represent the aggregate distribution of power
across space corresponding to the frequencies spanned by the bases
$\{b_{\nu}\}$. Observe that the sources are not explicitly present
in \eqref{eq:basis-expansion2}. Even if this model could have been
postulated directly for the cartography task at hand, the previous
discussion justifies the factorization of the $\Phi(\mathbf x,f)$
map per band in factors depending on each of the variables
$\mathbf{x}$ and $f$.

\section{Cooperative Spline-Based PSD Field Estimation}\label{sec:batch}

The sensing strategy will rely on the periodogram estimate
$\widehat \phi_{rn}(\tau)$ at a set of receiving (sampling)
locations
$\mathcal{X}:=\{\mathbf{x}_r\}_{r=1}^{N_r}\in\mathbb{R}^2$,
frequencies $\mathcal{F}:=\{f_n\}_{n=1}^N\in B$, and time-slots
$\{\tau\}_{\tau=1}^T$. In order to reduce the periodogram variance
 and mitigate fading effects, $\widehat \phi_{rn}(\tau)$ is
averaged across a window of $T$ time-slots~\cite{bazerque}, to
obtain
\begin{equation}\label{eq:intro-ewma}
 \varphi_{rn} :=\frac{1}{T}\sum_{\tau=1}^T
\widehat \phi_{rn}(\tau).
\end{equation}
Hence, the envisioned setup consists of $N_r$ receiving CRs, which
collaborate to construct the PSD map based on PSD observations
$\{\varphi_{rn}\}$. The bulk of processing is performed centrally
at a fusion center (FC), which is assumed to know the position
vectors $\mathcal{X}$ of all CRs, and the sensed tones in
$\mathcal{F}$. The FC receives over a dedicated control channel,
the vector of samples
$\bm\varphi_r:=[\varphi_{r1},\ldots,\varphi_{rN}]^{\prime}\in
\mathbb{R}^N$ taken by node $r$ for all $r=1,\ldots,N_r$.

While a BEM could be introduced for the spatial loss function
$l_s(\mathbf{x})$ as well \cite{bazerque}, the uncertainty on the
source locations and obstructions in the propagation medium may
render such a model imprecise. This will happen, e.g., when
shadowing is present. The alternative approach followed here
relies on estimating  the functions $g_\nu(\mathbf{x})$ based on
the data $\{\varphi_{rn}\}$. To capture the smooth portions of
$\Phi(\mathbf{x},f)$, the criterion for selecting
$g_\nu(\mathbf{x})$ will be regularized using a so termed
thin-plate penalty~\cite[p. 30]{wahba}. This penalty extends to
$\mathbb{R}^2$ the one-dimensional roughness regularization used
in smoothing spline models. Accordingly, functions
$\{g_\nu\}_{\nu=1}^{N_b}$ are estimated as
\begin{equation}\label{penalized_LS}
\{\hat
g_\nu\}_{\nu=1}^{N_b}:=\arg\hspace{-0.2cm}\min_{\{g_\nu\in\mathcal{S}\}}
\frac{1}{{N_rN}}\sum_{r=1}^{N_r}\sum_{n=1}^{N}
\left(\varphi_{rn}-\sum_{\nu=1}^{N_b}
g_\nu(\mathbf{x}_r)b_\nu(f_n)\right)^2+ \lambda
\sum_{\nu=1}^{N_b}\int_{\mathbb{R}^2} ||\nabla^2
g_\nu(\mathbf{x})||_F^2 d\mathbf{x}
\end{equation}
where  $||\nabla^2 g_\nu||_F$ denotes the Frobenius norm of the
Hessian of $g_\nu$.

The optimization is over $\mathcal{S}$, the space of Sobolev
functions, for which the penalty is well defined \cite[p.
85]{duchon}. The parameter $\lambda\geq 0$ controls the degree of
smoothing. Specifically, for $\lambda=0$ the estimates in
\eqref{penalized_LS} correspond to \textit{rough} functions
interpolating the data; while as $\lambda\to\infty$ the estimates
yield linear functions (cf. $\nabla^2 \hat
g_\nu(\mathbf{x})\equiv\mathbf{0}_{2\times 2}$). A smoothing
parameter in between these limiting values will be selected using
a leave-one-out cross-validation (CV) approach, as discussed
later.

\subsection{Thin-plate splines solution}\label{ssec:solution}

The optimization problem \eqref{penalized_LS} is variational in
nature, and in principle requires searching over the
infinite-dimensional functional space $\mathcal{S}$.
%
%
It turns out that \eqref{penalized_LS} admits  closed-form, finite
dimensional minimizers $\hat g_{\nu}(\mathbf{x})$, as presented in
the following proposition which provides a generalization of
standard thin-plate splines results; see e.g.,~\cite[p.31]{wahba},
to the multi-dimensional BEM (1).

\begin{proposition}\label{prop:thin-splines}
The estimates $\{\hat g_\nu\}_{\nu=1}^{N_b}$ in
\eqref{penalized_LS} are thin-plate splines expressible in
closed form as
\begin{equation}\label{eq:claim}
\hat g_\nu(\mathbf{x})=\sum_{r=1}^{N_r}\beta_{\nu r} K(||\mathbf{x}
-\mathbf{x}_r||_2)+\bm \alpha_{\nu 1}^{\prime}\mathbf{x}+\alpha_{\nu 0}
\end{equation}
where $K(\rho):=\rho^2\log(\rho)$, and $\bm\beta_\nu:=[\beta_{\nu
1},\ldots, \beta_{\nu N_r}]^{\prime}$ is constrained to the linear
subspace $\mathcal{B}:=\{\bm
\beta\in\mathbb{R}^{N_r}:\sum_{r=1}^{N_r}\beta_r=0,\sum_{r=1}^{N_r}\beta_r\mathbf{x}_r=\mathbf{0}_2,\:
\mathbf{x}_r\in\mathcal{X}\}$ for $\nu=1,\ldots,N_b$.
\end{proposition}

The proof of this proposition is given in Appendix A.

\begin{remark}\textbf{(Overlapping frequency
basis).}\label{remark:basis_pursuit} If the basis functions
$\{b_\nu(f)\}$ have finite supports which do not overlap, then
\eqref{penalized_LS} decouples per $g_\nu$,  and thus the results in
\cite{wahba,duchon} can be applied directly. The novelty of
Proposition \ref{prop:thin-splines} is that the basis functions with
spatial spline coefficients in \eqref{eq:basis-expansion2} are
allowed to be \emph{overlapping}. The implication of Proposition
\ref{prop:thin-splines} is finite parametrization of the PSD map
[cf. \eqref{eq:claim}]. This is particularly important for non-FDMA
based CR networks. In the forthcoming Section \ref{sec:BPSC}, an
overcomplete set $\{b_\nu\}$ is adopted in
\eqref{eq:basis-expansion2}, and overlapping bases naturally arise
therein.
\end{remark}

What is left to determine are the parameters
$\alpha:=[\alpha_{10},\bm \alpha_{11}^{\prime},\ldots,\alpha_{N_b
0}, \bm \alpha_{N_b 1}^{\prime}]^{\prime}\in\mathbb{R}^{3N_b}$,
and $\bm \beta:=[\bm \beta_1^{\prime},\ldots, \bm
\beta_{N_b}^{\prime}]^{\prime}\in\mathbb{R}^{N_rN_b}$ in
\eqref{eq:claim}. To this end, define the vector
$\bm\varphi:=[\varphi_{11},\ldots,\varphi_{1N},\ldots,\varphi_{N_r1},\ldots,\varphi_{N_rN}]^{\prime}\in\mathbb{R}^{{N_rN}}$
containing the network-wide data obtained at all frequencies in
$\mathcal F$. Three matrices are also introduced collecting the
regression inputs: i) $\mathbf{T}\in\mathbb{R}^{N_r\times 3}$ with
$r$th row $\mathbf{t}_r^{\prime}:=[1\:\mathbf{x}_r^{\prime}]$ for
$r=1,\ldots,N_r$ and $\mathbf{x}_r\in\mathcal{X}$; ii)
$\mathbf{B}\in\mathbb{R}^{N\times N_b}$ with $n$th row
$\mathbf{b}_n^{\prime}:=[b_{1}(f_n),\ldots, b_{N_b}(f_n)]$ for
$n=1,\ldots,N$; and iii) $\mathbf{K}\in\mathbb{R}^{N_r\times N_r}$
with $ij$-th entry
$[\mathbf{K}]_{ij}:=K(||\mathbf{x}_i-\mathbf{x}_j||)$ for
$\mathbf{x}_i,\mathbf{x}_j\in\mathcal{X}$. Consider also the QR
decompositions
of $\mathbf{T} =\left[\mathbf{Q}_1\:
\mathbf{Q}_2\right]\left[\mathbf{R}^\prime\
\mathbf{0}\right]^\prime$ and $\mathbf{B} =\left[\bm \Omega_1\:
\bm \Omega_2\right]\left[ \bm \Gamma^\prime\
\mathbf{0}\right]^\prime$.

Upon plugging \eqref{eq:claim} into \eqref{penalized_LS}, it is
shown in  Appendix B that the optimal $\{\bm\alpha,\bm\beta\}$
satisfy the following system of equations
\begin{align}
(\mathbf{B}\otimes
\mathbf{Q}_2^{\prime})\bm\varphi&=\left[(\mathbf{B}^{\prime}\mathbf{B}
\otimes \mathbf{Q}_2^\prime\mathbf{K}\mathbf{Q}_2)+{N_rN}\lambda
\mathbf{I}_{N_{b}(N_r-3)}\right]\hat{\bm\gamma}\label{system_eq1}\\
[\bm\Gamma \otimes \mathbf{R}]\hat{\bm\alpha}&=(\bm\Omega_1^{\prime}\otimes \mathbf{Q}_1^{\prime})
\bm\varphi - (\bm\Gamma \otimes \mathbf{Q}_1^{\prime}\mathbf{K}\mathbf{Q}_2)\hat{\bm\gamma}\label{system_eq2}\\
\hat{\bm\beta}&=(\mathbf{I}_{N_b}\otimes \mathbf{Q}_2)\hat{\bm\gamma}.\label{system_eq3}
\end{align}

Matrix $\mathbf{Q}_2^\prime\mathbf{K}\mathbf{Q}_2$ is positive
definite, and $\textrm{rank}(\bm\Gamma \otimes \mathbf{R})=
\textrm{rank}(\bm\Gamma)\textrm{rank}(\mathbf{R})$; see
e.g.,~\cite{minka}. It thus follows that
\eqref{system_eq1}-\eqref{system_eq2} admit a unique solution if
and only if $\bm\Gamma$ and $\mathbf{R}$ are invertible
(correspondingly, $\mathbf{B}$ and $\mathbf{T}$ have full column
rank). These conditions place practical constraints that should be
taken into account at the system design stage. Specifically,
$\mathbf{T}$ has full column rank if and only if the points in
$\mathcal{X}$, i.e., the CR locations, are not aligned.
Furthermore, $\mathbf{B}$ will have linearly independent columns
provided the basis functions $\{b_\nu(f)\}_{\nu=1}^{N_b}$ comprise
a linearly independent and complete set, i.e., $B\subseteq
\bigcup_{\nu}B_\nu$. Note that completeness precludes all
frequencies $\{f_n\}_{n=1}^{N}$ from falling outside the aggregate
support of the basis set, hence preventing undesired all-zero
columns in $\mathbf{B}$.

\begin{remark}\textbf{(Practicality of uniqueness conditions).}\label{remark:conditions}
\normalfont The condition on $\mathcal{X}$ does not introduce
an actual limitation as it can be easily satisfied in practice,
especially when the CRs are randomly deployed. Likewise, the
basis set is part of the system design, and can be chosen to
satisfy the conditions on $\mathbf{B}$. Nonetheless, these
conditions will be bypassed in Section \ref{sec:BPSC} by
allowing for an overcomplete set of functions $\{b_\nu\}$.
\end{remark}


The combined results in this section can be summarized in the
following steps constituting the spline-based spectrum cartography
algorithm, which amounts to estimating $\Phi(\mathbf{x},f)$:
\begin{description}
\item[\bf S1.]Given $\bm \varphi$, solve
    \eqref{system_eq1}-\eqref{system_eq3} for
    $\bm{\hat\alpha},\bm{\hat\beta}$, after selecting $\lambda$ as
    detailed in Appendix D.

\item[\bf S2.]Substitute $\bm{\hat\alpha}$ and $
    \bm{\hat\beta}$ into \eqref{eq:claim} to obtain $\{\hat
    g_\nu(\mathbf x)\}_{\nu=1}^{N_b}$.

\item[\bf S3.]Use $\{\hat
    g_\nu(\mathbf{x})\}_{\nu=1}^{N_b}$ in
    \eqref{eq:basis-expansion2} to estimate $\Phi(\mathbf
    x,f)$.
\end{description}


%
\subsection{PSD tracker}\label{ssec:online}

The real-time requirements on the sensing radios and the
convenience of an estimator that adapts to changes in the spectrum
map are the motivating reasons behind the  PSD tracker introduced
in this section. The spectrum map estimator will be henceforth
denoted by $\Phi(\mathbf{x},f,\tau)$, to make its time dependence
explicit.

Define the vector $\bm{\widehat
\phi}_n(\tau):=[\widehat\phi_{1n}(\tau),\ldots,\widehat\phi_{N_rn}(\tau)]^{\prime}$
of periodogram samples taken at frequency $f_n$ by all CRs, and
form the supervector $\bm{\widehat\phi}(\tau):=[\bm{\widehat
\phi}_1^{\prime}(\tau),\ldots,\bm{\widehat
\phi}_N^{\prime}(\tau)]^{\prime}\in\mathbb{R}^{N_rN}$. Per
time-slot $\tau=1,2,\ldots$, the periodogram
$\widehat{\bm\phi}(\tau)$ is averaged using the following adaptive
counterpart of \eqref{eq:intro-ewma}:
\begin{equation}\label{eq:recursive-phi}
\bm\varphi(\tau):=\sum_{\tau'=1}^{\tau}\delta^{\tau-\tau'}\widehat{\bm\phi}(\tau')=
\delta\bm\varphi(\tau-1)+\widehat{\bm\phi}(\tau)
\end{equation}
which implements an exponentially weighted moving average
operation with forgetting factor $\delta\in(0,1)$. For every
$\tau$, the online estimator $\Phi(\mathbf{x},f,\tau)$ is obtained
by plugging in \eqref{eq:basis-expansion2} the solution
$\{\hat{g}_{\nu}(\mathbf{x},\tau)\}_{\nu=1}^{N_b}$ of
\eqref{penalized_LS}, after replacing $\varphi_{rn}$ with
$\varphi_{rn}(\tau)$ [cf. the entries of the vector in
\eqref{eq:recursive-phi}]. In addition to mitigating fading
effects, this adaptive approach can track slowly time-varying PSDs
because the averaging in \eqref{eq:recursive-phi} exponentially
discards past data.

Suppose that per time-slot $\tau$, the FC receives raw
periodogram samples $\bm{\widehat\phi}(\tau)$ from the CRs in
order to update $\Phi(\mathbf{x},f,\tau)$. The results of
Section \ref{sec:batch} apply for every $\tau$, meaning that
$\{\hat{g}_{\nu}(\mathbf{x},\tau)\}_{\nu=1}^{N_b}$ are given by
\eqref{eq:claim}, while the optimum coefficients
$\{\hat{\bm\alpha}(\tau),\hat{\bm\beta}(\tau)\}$ are found
after solving \eqref{system_eq1}-\eqref{system_eq3}.
Capitalizing on \eqref{eq:recursive-phi}, straightforward
manipulations of \eqref{system_eq1}-\eqref{system_eq3} show
that $\{\hat{\bm\alpha}(\tau),\hat{\bm\beta}(\tau)\}$ are
recursively given for all $\tau\geq 1$ by
\begin{align}
\hat{\bm\beta}(\tau)&=\delta\hat{\bm\beta}(\tau-1)+
(\mathbf{I}_{N_b}\otimes
\mathbf{Q}_2)\mathbf{G}_1\widehat{\bm\phi}(\tau)\label{eq:update_beta}\\
\hat{\bm\alpha}(\tau)&=\delta\hat{\bm\alpha}(\tau-1)+
\mathbf{G}_2\widehat{\bm\phi}(\tau)\label{eq:update_alpha}
\end{align}
where the \textit{time-invariant} matrices $\mathbf{G}_1$ and
$\mathbf{G}_2$ are
\begin{align}
\mathbf{G}_1&:=\left[(\mathbf{B}^{\prime}\mathbf{B} \otimes
\mathbf{Q}_2^\prime\mathbf{K}\mathbf{Q}_2)+{N_rN}\lambda
\mathbf{I}_{N_{b}(N_r-3)}\right]^{-1}(\mathbf{B} \otimes
\mathbf{Q}_2^\prime)\nonumber\\
\mathbf{G}_2&:=[\bm\Gamma \otimes \mathbf{R}]^{-1}\left[
(\bm\Omega_1^{\prime}\otimes
\mathbf{Q}_1^{\prime})-(\bm\Gamma \otimes \mathbf{Q}_1^{\prime}\mathbf{K}
\mathbf{Q}_2)\mathbf{G}_1\right].\nonumber
\end{align}
Recursions \eqref{eq:update_beta}-\eqref{eq:update_alpha}
provide a means to update $\Phi(\mathbf{x},f,\tau)$
sequentially in time, by incorporating the newly acquired data
from the CRs in $\widehat{\bm\phi}(\tau)$. There is no need to
separately update $\bm\varphi(\tau)$ as in
\eqref{eq:recursive-phi}, yet the desired averaging takes
place. Furthermore, matrices $\mathbf{G}_1$ and $\mathbf{G}_2$
need to be computed only once, during the startup phase of the
network.

\section{Group-Lasso on Splines}\label{sec:BPSC}
An improved spline-based PSD estimator is developed in this
section to fit the unknown spatial functions
$\{g_\nu\}_{\nu=1}^{N_b}$ in the model
$\Phi(\mathbf{x},f)=\sum_{\nu=1}^{N_b}g_\nu(\mathbf{x})b_\nu(f)$,
with a large ($N_b\gg N_r N$), and a possibly  overcomplete set of
known basis functions $\{b_\nu\}_{\nu=1}^{N_b}$.
These models are particularly attractive when there is an inherent
uncertainty on the transmitters' parameters, such as central
frequency and bandwidth of the pulse shapers; or, e.g., the
roll-off factor when raised-cosine pulses are employed. In
particular, adaptive communication schemes rely on frequently
adjusting these parameters~\cite[Ch. 9]{goldsmith}. A sizeable
collection of bases to effectively accommodate most of the
possible cases provides the desirable robustness. Still, prior
knowledge available on the incumbent communication technologies
being sensed should be exploited to choose the most descriptive
classes of basis functions; e.g., a large set of raised-cosine
pulses. This knowledge justifies why known bases are selected to
describe frequency characteristics of the PSD map, while a
variational approach is preferred  to capture spatial
dependencies.

In this context, the envisioned estimation method should provide
the CRs with the capability of selecting a \emph{few} bases that
``better explain'' the actual transmitted signals.
As a result, most functions $g_\nu$ are expected to be
identically zero; hence, there is an inherent form of sparsity
present that can be exploited to improve estimation. The
rationale behind the proposed approach can be rooted in the
\textit{basis pursuit} principle, a term coined
in~\cite{basis_pursuit} for finding the most parsimonious
sparse signal expansion using an overcomplete basis set. A
major differentiating aspect however, is that while the sparse
coefficients in the basis expansions treated
in~\cite{basis_pursuit} are scalars, model
\eqref{eq:basis-expansion2} here entails bases weighted by
functions $g_\nu$.

%

The proposed approach to sparsity-aware spline-based field
estimation from the space-frequency power spectrum measurements
$\varphi_{rn}$ [cf. \eqref{eq:intro-ewma}], is to obtain  $\{\hat
g_\nu\}_{\nu=1}^{N_b}$ as
\begin{eqnarray}\label{penalized_LS_GL}
\{\hat
g_\nu\}_{\nu=1}^{N_b}:=\arg\hspace{-0.2cm}\min_{\{g_\nu\in\mathcal{S}\}}&&
\frac{1}{{N_rN}}\sum_{r=1}^{N_r}\sum_{n=1}^{N}
\left(\varphi_{rn}-\sum_{\nu=1}^{N_b}
g_\nu(\mathbf{x}_r)b_\nu(f_n)\right)^2 + \lambda
\sum_{\nu=1}^{N_b}\int_{\mathbb{R}^2} ||\nabla^2
g_\nu(\mathbf{x})||_F^2 d\mathbf{x}\nonumber\\
&&+\mu\sum_{\nu=1}^{N_b}\left\|\left[g_\nu(\mathbf{x}_1), \ldots,
g_\nu(\mathbf{x}_{N_r})\right]^{\prime}\right\|_2.
\end{eqnarray}
Relative to \eqref{penalized_LS}, the cost here is augmented with
an additional regularization term weighted by a tuning parameter
$\mu\geq 0$. Clearly, if $\mu=0$ then \eqref{penalized_LS_GL}
boils down to \eqref{penalized_LS}. To appreciate the role of  the
new penalty term, note that the minimization of
$\left\|\left[g_\nu(\mathbf{x}_1) ,\ldots,
g_\nu(\mathbf{x}_{N_r})\right]^{\prime}\right\|_2$ intuitively
shrinks all pointwise functional values
$\left\{g_\nu(\mathbf{x}_1) ,\ldots,
g_\nu(\mathbf{x}_{N_r})\right\}$ to zero for sufficiently large
$\mu$. Interestingly, it will be shown in the ensuing section that
this is enough to guarantee that $\hat g_\nu(\mathbf x)\equiv 0\
\forall \mathbf x$, for $\mu$ large enough.

\subsection{Estimation using the group-Lasso}\label{ssec:GL}
Consider the classical problem of linear regression; see,
e.g.~\cite[p. 11]{elements_of_statistics}, where a vector
$\mathbf{y}\in\mathbb{R}^n$ of observations is available, along
with a matrix $\mathbf{X}\in\mathbb{R}^{n\times p}$ of inputs.
The group
Lasso estimate for the vector of features
$\bm\zeta:=[\bm\zeta_1^\prime,\ldots,\bm\zeta_{N_b}^\prime]^\prime\in\mathbb{R}^p$
is defined as the solution
to~\cite{bakin_group_lasso,yuan_group_lasso}
\begin{equation}\label{group_lasso_orig}
\min_{\bm\zeta}\frac{1}{2}\left\|\mathbf{y}-\mathbf
X\bm\zeta\right\|_{2}^2 +\mu\sum_{\nu=1}^{N_b}\|\bm\zeta_\nu\|_2.
\end{equation}
 This criterion achieves
 model selection by retaining relevant factors  $\bm\zeta_\nu\in\mathbb R^{p/N_b}$
 in which the features are grouped.
 In other words, group-Lasso encourages sparsity at the factor
level, either by shrinking to zero all variables within a factor,
or by retaining them altogether depending on the value of the
tuning parameter $\mu\geq 0$. As $\mu$ is increased, more
sub-vector estimates $\bm\zeta_\nu$ become zero, and the
corresponding factors drop out of the model. It can be shown from
the Karush-Kuhn-Tucker optimality conditions that only for
$\mu\geq\mu_{\max}:=\max_i\|\mathbf{X}_i^\prime\mathbf{y}\|_2$ it
holds that $\bm\zeta_1=\ldots=\bm\zeta_{N_b}=\mathbf{0}_{p/N_b}$,
so that the values of interest are
$\mu\in[0,\mu_{\max}]$~\cite{danio_GL}.


The  connection between  \eqref{group_lasso_orig} and the
spline-based field estimator \eqref{penalized_LS_GL} builds on
Proposition \ref{prop:thin-splines}, which still holds in this
context. That is, even though criteria \eqref{penalized_LS} and
\eqref{penalized_LS_GL} purposely differ, their respective
solutions $\hat g_\nu(\mathbf{x})$ have the same form in
\eqref{eq:claim}. Indeed, the adaptation of the proof in Appendix
A to the new case is straightforward, since the additional penalty
term in \eqref{penalized_LS_GL} depends
on $g_\nu$ evaluated at the knots.  
The essential difference manifested by this penalty is revealed
when estimating the parameters $\bm\alpha $ and $\bm\beta$ in
\eqref{eq:claim}, as presented in the following proposition.

\begin{proposition}\label{prop:thin-splines-GL}
The spline-based field estimator   \eqref{penalized_LS_GL} is
equivalent to group-Lasso \eqref{group_lasso_orig}, under the
identities
\begin{align}\label{eq:identity_yX}
\mathbf y&:=\frac{1}{\sqrt{N_rN}}[\bm \varphi^\prime,\: \bm
0]^\prime,\quad \mathbf
X:=\frac{1}{\sqrt{N_rN}}\left[\begin{array}{c}
  \mathbf B\otimes \mathbf I_{N_r} \\
  \mathbf{I}_{N_b}\otimes\left\{\textrm{bdiag}(({N_rN}\lambda\mathbf{Q}_2^{\prime}\mathbf{K}\mathbf{Q}_2
)^{1/2},\mathbf{0})[\mathbf{KQ}_2\:\:\mathbf{T}]^{-1}\right\}
\end{array}\right]
\end{align}
with their respective solutions related by
\begin{align}\hat g_\nu(\mathbf{x})&=\sum_{r=1}^{N_r}\beta_{\nu r}
K(||\mathbf{x}-\mathbf{x}_r||_2)+\bm \alpha_{\nu
1}^{\prime}\mathbf{x}+\alpha_{\nu 0}\label{parametric_expansion_GL}\\
\left[\bm\beta_\nu^\prime,\bm\alpha_\nu^\prime\right]^\prime
&=\textrm{bdiag}(\mathbf{Q}_2,\mathbf{I}_3)[\mathbf{KQ}_2\:\:\mathbf{T}]^{-1}\hat{\bm\zeta}_\nu\label{change_of_variables_GL}
\end{align}
where $\bm\beta_\nu:=[\beta_{\nu1},\ldots,\beta_{\nu N_r}]^\prime$
and $\bm\alpha_\nu:=[\alpha_{\nu 0},\bm \alpha_{\nu
1}^\prime]^\prime$.
\end{proposition}
The factors $\{\bm\zeta_\nu\}_{\nu=1}^{N_b}$ in
\eqref{group_lasso_orig} are in one-to-one correspondence with the
vectors
$\{[{\bm\beta}_\nu^\prime,{\bm\alpha}_{\nu}^\prime]^\prime\}_{\nu=1}^{N_b}$
through the linear mapping \eqref{change_of_variables_GL}. This
implies that whenever a factor $\bm\zeta_\nu$ is dropped from the
linear regression model obtained after solving
\eqref{group_lasso_orig}, then $\hat g_\nu(\mathbf{x})\equiv 0$,
and the term corresponding to $b_\nu(f)$ does not contribute to
\eqref{eq:basis-expansion2}. Hence, by appropriately selecting the
value of $\mu$, criterion \eqref{penalized_LS_GL} has the
potential of retaining only the most significant terms in
$\Phi(\mathbf{x},f)=\sum_{\nu=1}^{N_b}g_\nu(\mathbf{x})b_\nu(f)$,
and thus yields parsimonious PSD map estimates. All in all, the
motivation behind the variational problem \eqref{penalized_LS_GL}
is now unravelled. The additional penalty term not present in
\eqref{penalized_LS} renders \eqref{penalized_LS_GL} equivalent to
a group-Lasso problem. This  enforces sparsity in the parameters
of the splines expansion for $\Phi(\mathbf x, f)$ at a factor
level, which is exactly what is needed to potentially null the
less descriptive functions $g_\nu$.

\begin{remark}\textbf{(Comparison with the PSD map estimator
in Section \ref{sec:batch}).}\label{remark:differences} The
sparsity-agnostic LS problem \eqref{penalized_LS} will not give rise
to identically zero vectors $\{\bm\alpha_\nu,\bm\beta_\nu\}$, for
any $\nu$. Even when $N_b$ is not large, a sparsity-aware estimator
will perform better if the underlying PSD is generated by a few
basis functions. This is expected since the out-of-band residual
error will increase when all basis functions enter the model
\eqref{eq:basis-expansion2}; see also~\cite{bazerque} for a related
assessment. What is more, when the number of bases is sufficiently
large ($N_b\gg {N_rN}$)  matrix $\mathbf{B}$ is fat, and  the
approach in Section \ref{sec:batch} is not applicable . On the other
hand, it is admittedly more complex computationally to solve
\eqref{group_lasso_orig} than the system of linear equations
\eqref{system_eq1}-\eqref{system_eq3}. Because
\eqref{penalized_LS_GL} is not a linear smoother, a leave-one-out
(bi-) CV approach to select the tuning parameters $\lambda$ and
$\mu$ does not enjoy the computational savings detailed in Appendix
D. $K$-fold CV can be utilized instead, with typical choices of
$K=5$ or $10$, as suggested in~\cite[p.
242]{elements_of_statistics}.
\end{remark}

The group-Lassoed splines-based approach to spectrum cartography
developed in this section can be summarized in the following steps
to estimate the global PSD map $\Phi(\mathbf{x},f)$:
\begin{description}
\item[\bf S1.]Given $\bm \varphi$ and utilizing any group
    Lasso solver, obtain
    $\hat{\bm\zeta}:=[\hat{\bm\zeta}_1^{\prime},\ldots,\hat{\bm\zeta}_{N_b}^{\prime}]^{\prime}$
    by solving \eqref{group_lasso_orig}.

\item[\bf S2.]Form the estimates
    $\bm{\hat\alpha},\bm{\hat\beta}$ using the change of
    variables
    $[\hat{\bm\beta}_\nu^\prime,\hat{\bm\alpha}_{\nu}^\prime]^\prime=
\textrm{bdiag}(\mathbf{Q}_2,\mathbf{I}_3)[\mathbf{KQ}_2\:\:\mathbf{T}]^{-1}\hat{\bm\zeta}_\nu$
for $\nu=1,\ldots,N_b$.

\item[\bf S3.]Substitute $\bm{\hat\alpha}$ and
    $\bm{\hat\beta}$ into \eqref{parametric_expansion_GL} to obtain
    $\{\hat g_\nu(\mathbf x)\}_{\nu=1}^{N_b}$.

\item[\bf S4.]Use $\{\hat
    g_\nu(\mathbf{x})\}_{\nu=1}^{N_b}$ in
    \eqref{eq:basis-expansion2} to estimate $\Phi(\mathbf
    x,f)$.
\end{description}

Implementing S1-S4 presumes that CRs communicate their local
PSD estimates to a fusion center, which uses their aggregation
in $\bm \varphi$ to estimate the field. But what if an FC is
not available for centrally running S1-S4? In certain cases,
forgoing with an FC is reasonable when the designer wishes to
avoid an isolated point of failure, or, aims at a network
topology which scales well with an increasing number of CRs
based on power considerations (CRs located far away from the FC
will drain their batteries more to reach the FC). These reasons
motivate well a fully distributed counterpart of S1-S4, which
is pursued next.

\section{Distributed Group-Lasso for In-Network Spectrum
Cartography}\label{sec:dglasso}

Consider $N_r$ networked CRs that are capable of sensing the
ambient RF spectrum, performing some local computations, as well
as exchanging messages among neighbors via dedicated control
channels. In lieu of a fusion center, the CR network is naturally
modeled as an undirected graph $\calG(\calR,\calE)$, where the
vertex set $\calR:=\{1,\ldots,N_r\}$ corresponds to the sensing
radios, and the edges in $\calE$ represent pairs of CRs that can
communicate. Radio $r\in\calR$ communicates with its single-hop
neighbors in $\calN_r$, and the size of the neighborhood is
denoted by $|\calN_r|$. The locations
$\{\mathbf{x}_r\}_{r=1}^{N_r}:=\mathcal{X}$ of the sensing radios
are assumed known to the CR network. To ensure that the measured
data from an arbitrary CR can eventually percolate throughout the
entire network, it is assumed that the graph $\calG$ is
\emph{connected}; i.e., there exists a (possibly) multi-hop
communication path connecting any two CRs.

For the purpose of estimating an unknown vector
$\bm\zeta:=\left[\bm\zeta_1^\prime,\ldots,\bm\zeta_{N_b}^\prime\right]^\prime\in\mathbb{R}^p$,
each radio $r\in\calR$ has available a local vector of
observations $\mathbf{y}_r\in\mathbb{R}^{n_r}$ as well as its own
matrix of inputs $\mathbf X_r \in\mathbb{R}^{n_r\times p}$. Radios
collaborate to form the wanted group-Lasso estimator
\eqref{group_lasso_orig} in a distributed fashion, using
\begin{equation}\label{group_lasso_dist}
\hat{\bm\zeta}_{\textrm{glasso}}=
\mbox{arg}\:\min_{\bm\zeta}\frac{1}{2}\sum_{r=1}^{N_r}\left\|\mathbf{y}_r-\mathbf{X}_{r}\bm\zeta
\right\|_2^{2}+ \mu\sum_{\nu=1}^{N_b}\|\bm\zeta_\nu\|_2
\end{equation}
where $\bby:=[\bby_1^{\prime},\ldots,\bby_{N_r}^{\prime}]^{\prime}
\in \mathbb{R}^{n\times 1}$ with $n:=\sum_{r=1}^{N_r}n_r$, and
$\bbX:=[\bbX_1^{\prime},\ldots,\bbX_{N_r}^{\prime}]^{\prime}\in\mathbb{R}^{n\times
p}$. The motivation behind developing a distributed solver of
\eqref{group_lasso_dist} is to tackle \eqref{penalized_LS_GL}
based on in-network processing of the local observations
$\bm\varphi_r:=[\varphi_{r1},\ldots,\varphi_{r N}]^{\prime}$
available per radio [cf. \eqref{eq:intro-ewma}]. Indeed, it
readily follows that \eqref{group_lasso_dist}  can be used instead
of \eqref{group_lasso_orig}  in Proposition
\ref{prop:thin-splines-GL} when
\begin{equation*}
\mathbf{y}_r:=\frac{1}{\sqrt{N_rN}}\left[\begin{array}{c}\bm\varphi_r\\
\mathbf{0}\end{array}\right],\quad \mathbf{X_r}:=\frac{1}{\sqrt{N_rN}}\left[\begin{array}{c}\mathbf{B}\otimes\mathbf{e}_{N_r,r}^{\prime}\\
\mathbf{I}_{N_b}\otimes\left[\textrm{bdiag}((N_rN\lambda\mathbf{Q}_2^{\prime}\mathbf{K}\mathbf{Q}_2
)^{1/2},\mathbf{0})[\mathbf{KQ}_2\:\:\mathbf{T}]^{-1}\right]\end{array}\right],\quad
r\in\calR
\end{equation*}
corresponding to the identifications $n_r=N$ $\forall r\in\calR$,
$p=N_b N_r$. Note that because the locations $\{\mathbf x_r\}$ are
assumed known to the entire network, CR $r$ can form matrices
$\mathbf T$, $\mathbf K$, and thus, the local regression matrix
$\mathbf X_r$.
\subsection{Consensus-based reformulation of the group-Lasso}
\label{ssec:reformulation}

To distribute the cost in \eqref{group_lasso_dist}, replace the
\textit{global} variable $\bm \zeta$ which couples the per-agent
summands with \textit{local} variables $\{\bm
\zeta_{r}\}_{r=1}^{N_r}$ representing candidate estimates of $\bm
\zeta$ per sensing radio. It is now possible to reformulate
\eqref{group_lasso_dist} as a convex \emph{constrained}
minimization problem
\begin{align}\label{constrmin_0}
\left\{\hat{\bm\zeta}_{r}\right\}_{r=1}^{N_r}=&\:
\mbox{arg}\:\min_{\{\bm\zeta_{r}\}}\frac{1}{2}\sum_{r=1}^{N_r}\left[\left\|\mathbf{y}_r-\mathbf{X}_{r}\bm\zeta_{r}\right\|_2^{2}+
\frac{2\mu}{N_r}\sum_{\nu=1}^{N_b}\|\bm\zeta_{r\nu}\|_2\right]\\
\nonumber&\begin{array}{ccl}\mbox{ s.
t.}&{\:}{\:}&\bm\zeta_r=\bm\zeta_{r^{\prime}},{\:} r\in\calR,{\:}
r^{\prime}\in\calN_{r},\ \bm
\zeta_r:=\left[\bm\zeta_{r1}^\prime,\ldots,\bm
\zeta_{rN_b}^\prime\right]^\prime.
\end{array}
\end{align}
The equality constraints directly effect local agreement across
each CR's neighborhood. Since the communication graph
$\mathcal{G}$ is  assumed connected, these constraints also
ensure \textit{global} consensus a fortiori, meaning that
$\bm\zeta_r=\bm\zeta_{r^\prime},$ $\forall r,r^\prime\in\calR$.
Indeed, let $P(r,r^\prime):r,r_1,r_2,\ldots,r_n,r^\prime$
denote a path on $\mathcal{G}$ that joins an arbitrary pair of
CRs $(r,\rpr)$. Because contiguous radios in the path are
neighbors by definition, the corresponding chain of equalities
$\bm\zeta_r=\bm\zeta_{r_1}=\bm\zeta_{r_2}=\ldots=\bm\zeta_{r_n}=\bm\zeta_{\rpr}$
dictated by the constraints in \eqref{constrmin_0} imply
$\bm\zeta_r=\bm\zeta_{\rpr}$, as desired. Thus, the constraints
can be eliminated by replacing all the $\{\bm\zeta_r\}$ with a
common $\bm\zeta$, in which case the cost in
\eqref{constrmin_0} reduces to the one in
\eqref{group_lasso_dist}. This argument establishes the
following result.

\begin{lemma}\label{prop:equiv_consensus}
If $\mathcal{G}$ is a connected graph, \eqref{group_lasso_dist}
and \eqref{constrmin_0} are equivalent optimization problems, in
the sense that
$\hat{\bm\zeta}_{\textrm{glasso}}=\hat{\bm\zeta}_{r},{\:}\forall{\:}r\in\mathcal{R}.$
\end{lemma}

Problem \eqref{constrmin_0} will be  modified further for the
purpose of reducing the computational complexity of the
resulting algorithm. To this end, for a given $\mathbf a\in
\mathbb R^p$ consider the problem
\begin{align}
\min_{\bm \zeta} &\frac{1}{2}||\bm \zeta||_2^2 -\mathbf a^\prime
\bm \zeta + \mu
\sum_{\nu=1}^{N_b}\|\bm\zeta_{\nu}\|_2,\hspace{1cm}\bm\zeta:=[\bm
\zeta_{1}^\prime,\ldots,\bm\zeta_{N_b}^\prime]^\prime\label{glasso_tonto}
\end{align}
and notice that it is separable in the $N_b$ subproblems
\begin{align}
\min_{\bm \zeta_{\nu}} &\frac{1}{2}||\bm \zeta_{\nu}||_2^2
-\mathbf a_\nu^\prime \bm \zeta_{\nu} +
\mu\|\bm\zeta_{\nu}\|_2,\hspace{1cm}\mathbf a:=[\mathbf
a_1^\prime,\ldots,\mathbf
a_{N_b}^\prime]^\prime\label{glasso_super_tonto}.
\end{align}
Interestingly, each of these subproblems admits a closed-form solution as given  in the following lemma.

\begin{lemma}\label{prop:glasso_closed_form}
The minimizer $\bm\zeta^\star_{\nu}$ of \eqref{glasso_super_tonto}
is obtained via the vector
 soft-thresholding operator $\mathcal T_\mu(\cdot)$ defined by
\begin{align}
\bm\zeta_{\nu}^\star=\mathcal T_\mu(\mathbf{a}_\nu):=\frac{\mathbf
a_\nu}{\|\mathbf a_\nu\|_2}\left(\|\mathbf
a_\nu\|_2-\mu\right)_+\label{solucion_tonta}
\end{align}
where $\left(\cdot\right)_+:=\max\{\cdot,0\}$ .
\end{lemma}

Problem \eqref{glasso_tonto} is an instance of the group-Lasso
\eqref{group_lasso_orig} when $\mathbf X^\prime\mathbf X=\mathbf
I_p$, and $\mathbf a:=\mathbf X^\prime\bby$. As such, result
\eqref{solucion_tonta} can be viewed as a particular case of the
operators in \cite{puig_hero} and \cite{wright_novak_figueredo}.
However it is worth to prove Lemma \ref{prop:glasso_closed_form}
directly, since in this case the special form of
\eqref{glasso_super_tonto} renders the proof neat in its
simplicity.

\begin{proof}
It will be argued that the solver of \eqref{glasso_super_tonto}
takes the form
 $\bm\zeta_{\nu}^\star=t \mathbf{a}_\nu$ for some scalar $t\geq 0$. This is because among all  $\bm \zeta_{\nu}$
  with the same $\ell_2$-norm, the Cauchy-Schwarz inequality implies that the maximizer of
  $\mathbf a_\nu^\prime\bm \zeta_{\nu}$ is colinear  with (and  in the same direction of)   $\mathbf a_\nu$.
  Substituting $\bm\zeta_{\nu}=t \mathbf a_\nu$ into \eqref{glasso_super_tonto} renders the  problem scalar in $t\geq 0$, with
   solution $t^\star =  \left(\|\mathbf a_\nu\|-\mu\right)_+/\left({2\|\mathbf a_\nu\|}\right)$, which completes the proof.
\end{proof}

In order to take advantage of  Lemma
\ref{prop:glasso_closed_form}, auxiliary variables $\bm \gamma_r,\
r=1,\ldots,N_r$ are
 introduced as copies of $\bm\zeta_r$. Upon introducing appropriate constraints  $\bm \gamma_r=\bm\zeta_r$ that guarantee the equivalence of the formulations along the lines of Lemma \ref{prop:equiv_consensus},
 problem \eqref{constrmin_0} can be recast as
\begin{align}\label{constrmin_1}
\left\{\hat{\bm\zeta}_{r}\right\}_{r=1}^{N_r}=&\hspace{0.3cm}
\mbox{arg}\hspace{-0.4cm}\min_{\{\bm\zeta_{r},\bm \gamma_r,\bm
\gamma_r^{r^\prime}\}}\frac{1}{2}\sum_{r=1}^{N_r}\left[\left\|\mathbf{y}_r-\mathbf{X}_{r}\bm\gamma_{r}\right\|_2^{2}+
\frac{2\mu}{N_r}\sum_{\nu=1}^{N_b}\|\bm\zeta_{r\nu}\|_2\right]\\
\nonumber&\begin{array}{ccl}\mbox{ s.
to}&{\:}{\:}&\bm\zeta_r=\bm\gamma_{r}^{r^\prime}=\bm\zeta_{r^{\prime}},{\:}
r\in\calR,{\:}
r^{\prime}\in\calN_{r}\\
&&\bm \gamma_r=\bm\zeta_r,\ r\in\calR .\end{array}
\end{align}
The  dummy  variables  $\bm \gamma_r^{r^\prime}$ are inserted for
technical reasons that will become apparent in the ensuing
section, and will be eventually eliminated.

\subsection{Distributed group-Lasso algorithm}
\label{ssec:dglasso}

The distributed group-Lasso algorithm is constructed by optimizing
\eqref{constrmin_1} using the alternating direction method of
multipliers (AD-MoM) \cite{Bertsekas_Book_Distr}. In this
direction, associate Lagrange multipliers $\bbv
_r,\bar{\bbv}_{r}^{r^\prime}$ and  $\breve{\bbv}_{r}^{r^\prime}$
with the constraints  $\bm \gamma_r=\bm\zeta_r$,
$\bm\zeta_{r^\prime}=\bm\gamma_{r}^{r^\prime}$ and
$\bm\zeta_r=\bm\gamma_{r}^{r^\prime}$,
 respectively, and
consider the augmented Lagrangian with parameter $c>0$
\begin{align}\label{Aug_Lagr_1}
\ccalL_{c}\left[\{\bm\zeta_r\},\bm\gamma, \calbbv\right]=&
\frac{1}{2}\sum_{r=1}^{N_r}\left[\left\|\mathbf{y}_r-\mathbf{X}_{r}\bm\gamma_{r}\right\|_2^{2}
+\frac{2\mu}{N_r}\sum_{\nu=1}^{N_b}\|\bm\zeta_{r\nu}\|_2\right]
+\sum_{r=1}^{N_r}\bbv_r^\prime(\bm\zeta_r-\bm\gamma_r)
+\frac{c}{2}\sum_{r=1}^{N_r}
\|\bm\zeta_r-\bm\gamma_r\|_2^{2}\nonumber\\
&\hspace{-1.8cm}+\sum_{r=1}^{N_r}\sum_{r^{\prime}\in\calN_{r}}\left[(\breve{\bbv}_{r}^{r^\prime})^\prime
(\bm\zeta_{r}-{\bm\gamma}_r^{r^\prime})+(\bar{\bbv}_r^{r^\prime})^\prime
(\bm\zeta_{r^\prime}-{\bm\gamma}_{r}^{r^\prime})\right]
+\frac{c}{2}\sum_{r=1}^{N_r}\sum_{r^\prime\in\calN_r}
\left[\|\bm\zeta_{r}-{\bm\gamma}_r^{r^\prime}\|_2^{2}+
\|\bm\zeta_{r^\prime}-{\bm\gamma}_{r}^{r^\prime}\|_2^2\right]
\end{align}
where for notational convenience we group   the variables $\bm\gamma:=\{\bm\gamma_r,\{{\bm\gamma}_r^{r^\prime}\}_{r^\prime\in\calN_r}
\}_{r\in\calR}$, and multipliers\\
$\calbbv:=\{\bbv_r,\{\breve{\bbv}_{r}^{r^\prime}\}_{r^\prime\in\calN_r},\{\bar{\bbv}_{r}^{r^\prime}\}_
{r^\prime\in\calN_r}\}_{r\in\calR}$.

Application of the AD-MoM to the problem at hand consists of a cycle
of $\ccalL_c$ minimizations in a block-coordinate fashion w.r.t.
$\{\bm \zeta_r\}$ firstly, and $\bm \gamma$ secondly, together with
an update of the multipliers per iteration $k=0,1,2,\ldots$.
Deferring the details to Appendix E, the four main properties of
this procedure that are instrumental to the resulting algorithm can
be highlighted as follows.
\begin{enumerate}
\item[i)]  Thanks to  the introduction of the local copies $\bm
\zeta _r$ and the dummy variables $ \bm\gamma_r^{r^\prime}$,  the
minimizations of $\ccalL_c$ w.r.t. both $\{\bm \zeta_r\}$ and $\bm
\gamma$ decouple per CR  $r$, thus enabling distribution of the
algorithm. Moreover, the constraints in \eqref{constrmin_1}
involve variables of neighboring CRs only, which allows the
required communications to be local within each CR's neighborhood.

\item[ii)] Introduction of the variables $\bm\gamma_r$ separates
the quadratic cost $\|\bby_r-\mathbf X_r\bm \gamma_r\|_2^2$ from
the group-Lasso penalty $\sum_{\nu=1}^{N_b}\|\bm\zeta_{r\nu}\|_2$.
As a result, minimization of \eqref{Aug_Lagr_1} w.r.t.
$\bm\zeta_r$ takes the form of \eqref{glasso_tonto}, which admits
a closed-form solution via the vector soft-thresholding operator
$\mathcal T_\mu(\cdot)$ in Lemma \ref{prop:glasso_closed_form}.

\item[iii)] Minimization of \eqref{Aug_Lagr_1} w.r.t. $\bm\gamma$
consists of an unconstrained quadratic  problem, which can also be
solved in closed form. In particular, the optimal $\bm
\gamma_{r}^{r^\prime}$ at iteration $k$ takes the value $\bm
\gamma_{r}^{r^\prime}(k)=\left(\bm \zeta_{r}(k)+\bm
\zeta_{r^\prime}(k)\right)/2$, and thus can be eliminated.

\item[iv)] It turns out that it is not necessary to carry
    out updates of the Lagrange multipliers $\{\bar{
    \bbv}_r^{r^\prime},\ \breve{
    \bbv}_r^{r^\prime}\}_{r^\prime\in \mathcal N_r}$
    separately, but only of their sums which are henceforth
    denoted by $\bbp_r:=\sum_{r^\prime\in
    \ccalN_r}\left(\bar{\bbv}_r^{r^\prime}+\breve{\bbv}_r^{r^\prime}\right)$.
    Hence, there is one price $\bbp_r$ per CR
    $r=1,\ldots,N_r$, which can be updated locally.
\end{enumerate}

Building on these four features, it is established in Appendix E
that the proposed AD-MoM scheme boils down to four parallel
recursions run locally per CR:
\begin{align}
\label{pupdate_1}\bbp_{r}(k)&=\bbp_{r}(k-1)+c\sum_{r^\prime\in\calN_r}[\bm\zeta_r(k)-\bm\zeta_{r^\prime}(k)]\\
\label{update_1}\bbv_{r}(k)&=\bbv_{r}(k-1)+c[\bm\zeta_{r}(k)-\bm\gamma_r(k)]\\
\label{betaupdate_1}\bm\zeta_{r\nu}(k+1)&=\frac{\mathcal
T_{\mu}\left(N_r\left(c\bm\gamma_{r\nu}(k)+c\sum_{r^\prime\in\calN_r}\left[
\bm\zeta_{r\nu}(k)+\bm\zeta_{r^\prime\nu}(k)\right]-\mathbf{p}_{r\nu}(k)-
\mathbf{v}_{r\nu}(k)\right)\right)}{cN_r(2|\calN_r|+1)},\:\: \nu=1,\ldots,N_b \\
\label{gammaupdate_1}\bm\gamma_r(k+1)&=\left[c\mathbf{I}_p+\mathbf{X}_r^\prime
\mathbf{X}_r\right]^{-1}\left(\mathbf{X}_r^\prime\mathbf{y}_r
+c\bm\zeta_r(k+1)+\mathbf{v}_r(k)\right).
\end{align}
Recursions \eqref{pupdate_1}-\eqref{gammaupdate_1} comprise the
novel DGLasso algorithm, tabulated as Algorithm
\ref{DGLasso_algorithm_table}.

\begin{algorithm}[t]
\caption{: DGLasso} \small{
\begin{algorithmic}
    \STATE All radios $r\in\calR$ initialize
    $\{\bm\zeta_r(0),\bm\gamma_r(0),\bbp_r(-1),\bbv_r(-1)\}$  to zero, and locally
    run:
    \FOR {$k=0,1$,$\ldots$}
        \STATE Transmit $\bm\zeta_{r}(k)$ to neighbors in $\calN_r$.
        \STATE Update $\bbp_{r}(k)$ via $\bbp_{r}(k)=\bbp_{r}(k-1)+c\sum_{r^\prime\in\calN_r}[\bm\zeta_r(k)-\bm\zeta_{r^\prime}(k)]$.
        \STATE Update $\bbv_r(k)$ via $\bbv_{r}(k)=\bbv_{r}(k-1)+c[\bm\zeta_{r}(k)-\bm\gamma_r(k)]$.
        \STATE Update $\bm\zeta_{r}(k+1)$ using \eqref{betaupdate_1}.
        \STATE Update $\bm\gamma_{r}(k+1)$ using \eqref{gammaupdate_1}.
    \ENDFOR
\end{algorithmic}}
\label{DGLasso_algorithm_table}
\end{algorithm}

The algorithm entails the following steps. During iteration $k+1$,
CR $r$ receives the local estimates
$\{\bm\zeta_{r^\prime}(k)\}_{r^\prime\in\calN_r}$ from the
neighboring CRs and plugs them into \eqref{pupdate_1} to evaluate
the dual price vector $\bbp_r(k)$. The new multiplier $\bbv_r(k)$
is then obtained using the locally available vectors
$\{\bm\gamma_r(k),\bm\zeta_r(k)\}$. Subsequently,  vectors
$\{\bbp_r(k),\bbv_r(k)\}$ are jointly used along with
$\{\bm\zeta_{r^\prime}(k)\}_{r^\prime\in\calN_r}$ to obtain
$\bm\zeta_r(k+1)$ via $N_b$ parallel vector soft-thresholding
operations $\mathcal T_\mu(\cdot)$ as in \eqref{solucion_tonta}.
Finally, the updated $\bm\gamma_r(k+1)$ is obtained from
\eqref{gammaupdate_1}, and requires the previously updated
quantities along with the  vector of local observations $\bby_r$
and regression matrix $\mathbf{X}_r$. The $(k+1)$st iteration is
concluded after CR $r$ broadcasts $\bm\zeta_r(k+1)$ to its
neighbors. Even if an arbitrary initialization is allowed, the
sparse nature of the estimator sought suggests the all-zero
vectors as a natural choice. Three additional remarks are now in
order.

\begin{remark}\textbf{(Distributed  Lasso
algorithm as a special case).} \normalfont When $N_b=p$ and
there are as many groups as entries of $\bm \zeta$, then the
sum $\sum_{\nu=1}^{N_b}\|\bm\zeta_\nu\|$ becomes the
$\ell_1$-norm of $\bm \zeta$, and group-Lasso reduces to Lasso.
In this case, DGLasso offers a distributed algorithm to solve
Lasso that  coincides with the one in \cite{bazerque_dlasso}.
\end{remark}

\begin{remark}\textbf{(Centralized Group-Lasso algorithm as a special case).}
\normalfont For $N_r=1$, the network consists of a single CR. In
this case, DGLasso yields a novel algorithm for the centralized
group-Lasso estimator \eqref{group_lasso_dist}, which is specified
as Algorithm \ref{GLasso_algorithm_table}. Notice that  the
thresholding operator $\mathcal T_\mu$ in GLasso sets the entire
sub-vector $\bm \zeta_\nu(k+1)$ to zero whenever
 $\|c\bm\gamma_\nu(k)-\bbv_\nu(k)\|_2$ does not
exceed $\mu$, in par with the group-sparsifying property of
group-Lasso. Different from \cite{yuan_group_lasso}, GLasso can
handle a general (not orthonormal) regression matrix $\mathbf X$.
Compared to the block-coordinate algorithm of \cite{puig_hero},
GLasso does not require an inner Newton-Raphson recursion per
iteration. If in addition $N_b=p$, then GLasso yields the Lasso
estimator.
\end{remark}

\begin{algorithm}[t]
\caption{: GLasso} \small{
\begin{algorithmic}
    \STATE Initialize $\{\bm\zeta(0),\bm \gamma(0),\bbv(-1)\}$ to zero,  and
    run:
    \FOR {$k=0,1$,$\ldots$}
        \STATE Update $\bbv(k)=\bbv(k-1)+c[\bm\zeta(k)-\bm\gamma(k)]$.
        \STATE Update $\bm\zeta_\nu(k+1)=(1/c)\mathcal T_\mu\left(c\bm\gamma_\nu(k)-\bbv_\nu(k)\right),\ \nu=1,\ldots,N_b$.\label{betaupdate_centralized}
        \STATE Update $\bm\gamma(k+1)=\left[c\mathbf I_p+ \mathbf X^\prime\mathbf X\right]^{-1}\left(\mathbf X^\prime\bby+c\bm \zeta(k+1)+\bbv(k)\right)$.
    \ENDFOR
\end{algorithmic}}
\label{GLasso_algorithm_table}
\end{algorithm}

\begin{remark}\textbf{(Computational load balancing).}\label{remark:load_balance}
\normalfont Update \eqref{gammaupdate_1} involves inversion of the
$p\times p$ matrix
$c\mathbf{I}_p+\mathbf{X}_r^{\prime}\mathbf{X}_r$, that may be
computationally demanding for sufficiently large $p$. Fortunately,
this operation can be carried out offline before running the
algorithm. More importantly, the matrix inversion lemma can be
invoked to obtain
$\left[c\mathbf{I}_p+\mathbf{X}_r^{\prime}\mathbf{X}_r\right]^{-1}=c^{-1}
\left[\mathbf{I}_p-\mathbf{X}_r^{\prime}\left(c\mathbf{I}_{n_r}+\mathbf{X}_r\mathbf{X}_r^{\prime}\right)^{-1}\mathbf{X}_r\right]$.
In this new form, the dimensionality of the matrix to invert
becomes $n_r\times n_r$, where $n_r$ is the number of locally
acquired data. For highly underdetermined cases $(n_r\ll p)$,
(D)GLasso enjoys considerable computational savings through the
aforementioned matrix inversion identity. One also recognizes that
the distributed operation parallelizes the numerical computation
across CRs: if GLasso is run at a central unit with all
network-wide data available centrally, then the matrix to invert
has dimension $n=\sum_{r\in\mathcal{R}}n_r$, which increases
linearly with the network size $N_r$. Beyond a networked scenario,
DGLasso provides an attractive alternative for computational load
balancing in contemporary multi-processor architectures.
\end{remark}

To close this section, it is useful to  mention that
convergence of Algorithm 1, and thus of Algorithm 2 as well, is
ensured by the convergence of the AD-MoM
\cite{Bertsekas_Book_Distr}. This result is formally stated
next.

\begin{proposition}
\label{prop:convergence_DGL} Let $\calG$ be a connected graph, and
consider recursions \eqref{pupdate_1}-\eqref{gammaupdate_1} that
comprise the DGLasso algorithm. Then, for any value of the
step-size $c>0$, the iterates $\bm\zeta_r(k)$ converge to the
group-Lasso solution [cf. \eqref{group_lasso_dist}] as
$k\to\infty$, i.e.,
\begin{equation}
\lim_{k\to\infty}\bm\zeta_r(k)=\hat{\bm\zeta}_{\textrm{glasso}},{\:}\forall{\:}r\in\calR.\label{consensus_attained}
\end{equation}
\end{proposition}
 In words, all local estimates $\bm\zeta_r(k)$ achieve consensus
asymptotically, converging to a common vector that coincides with
the desired estimator $\hat{\bm\zeta}_{\textrm{glasso}}$.
Formally, if the number of parameters $p$ exceeds the number of
data $n$, then a unique solution of \eqref{group_lasso_orig} is
not guaranteed for a general design matrix $\bbX$. Proposition
\ref{prop:convergence_DGL} remains valid however, if the
right-hand side of \eqref{consensus_attained} is replaced by the
set of minima; that is,
\begin{equation*}\lim_{k\to\infty}\bm\zeta_r(k) \in
\mbox{arg}\:\min_{\bm\zeta}\frac{1}{N_r}\sum_{r=1}^{N_r}\left\|\mathbf{y}_r-\mathbf{X}_{r}\bm\zeta
\right\|_2^{2}+
\mu\sum_{\nu=1}^{N_b}\|\bm\zeta_\nu\|_2.\end{equation*}

\section{Numerical tests}\label{sec:sims}

\begin{figure}[h]
\begin{minipage}[b]{.48\linewidth}
  \centering
  \centerline{\includegraphics[width=6cm]{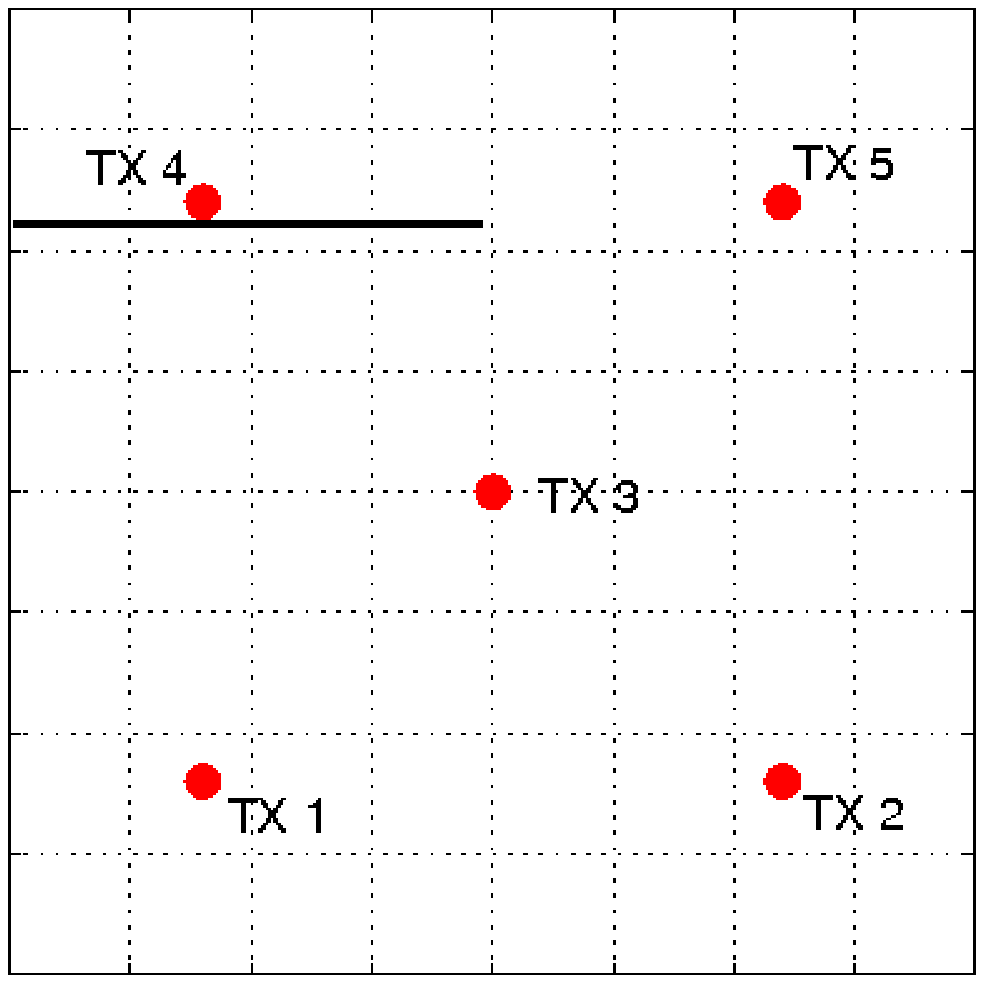}}
 \vspace{-0.8cm}
\medskip
\end{minipage}
\hfill
\begin{minipage}[b]{0.48\linewidth}
  \centering
  \centerline{\includegraphics[width=8cm]{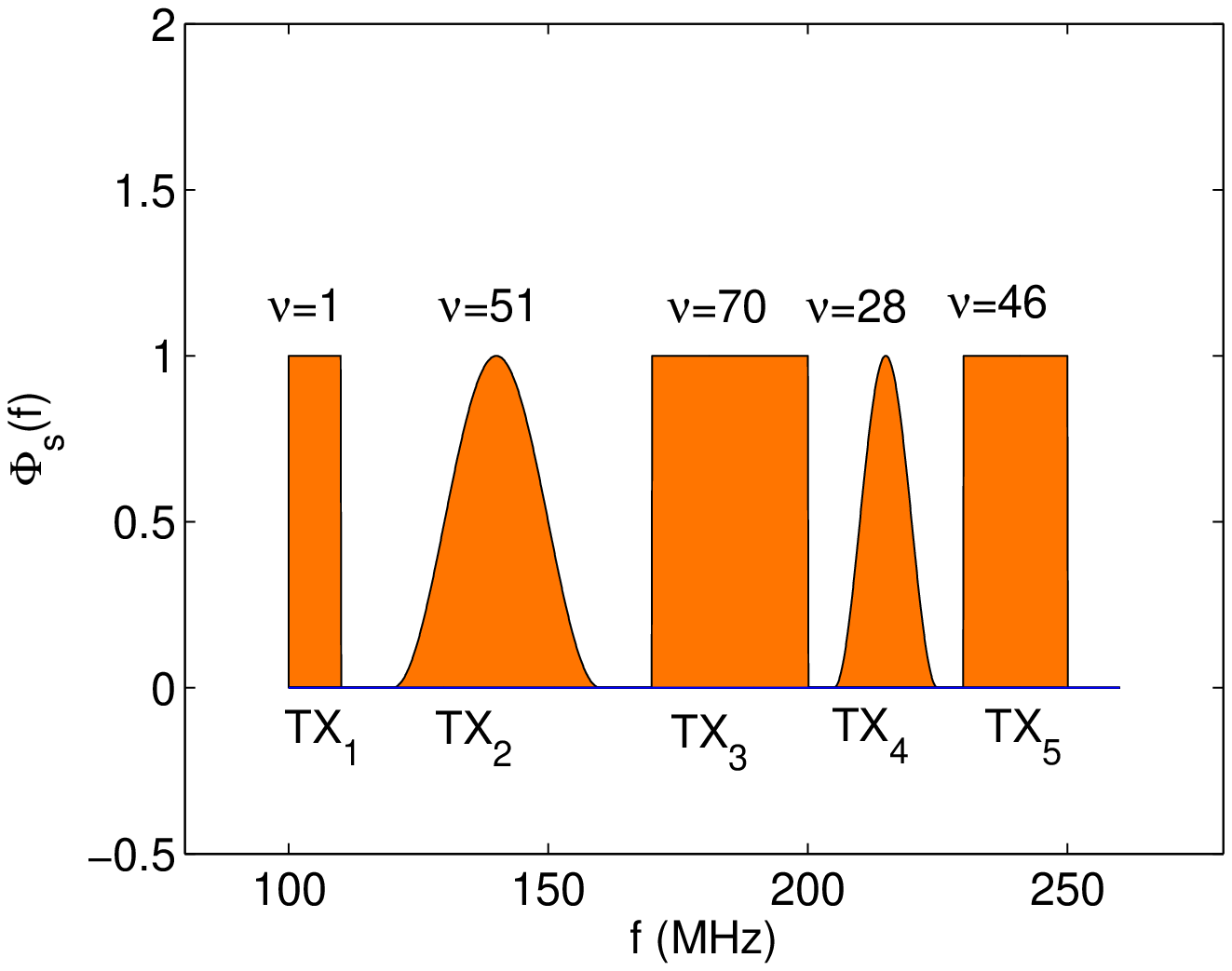}}
 \vspace{-0.8cm}
\medskip
\end{minipage}

\caption{(left) Position of sources and obstructing wall;
(right) PSD generated by the active transmitters.}
\label{fig:dado_bases}
\end{figure}

Consider a set of $N_r=100$ CRs uniformly distributed in an
area of $1{\rm Km}^2$, cooperating to estimate the PSD map
generated by $N_s=5$ licensed users (sources) located as in
Fig. \ref{fig:dado_bases} (left). The five transmitted signals
are raised cosine pulses with roll-off factors
$\rho\in\{0,1\}$, and bandwidths $W\in\{10, 20, 30\}$ MHz. They
share the frequency band $B=[100,260]$ MHz with spectra
centered at frequencies $f_c=105,\ 140,\ 185,\ 215,$ and $240$
MHz, respectively. Fig. \ref{fig:dado_bases} (right) depicts
the PSD generated by the active transmitters.

The PSD generated by source $s$ experiences fading and shadowing
effects in its propagation from $\mathbf{x}_s$ to any location
$\mathbf{x}$, where it can be measured in the presence of noise. A
6-tap Rayleigh model is adopted for the multipath channel
$H_s(f,\tau,\mathbf{x})$ between $\mathbf{x}_s$ and $\mathbf{x}$,
whose expected gain adheres to the path-loss law
$E(|H_s|^2)=\exp(-||\mathbf{x}_s-\mathbf{x}||_2^2/\Delta^2)$, with
$\Delta=0.8$. A deterministic shadowing effect is generated by a
$18$m-high and $500$m-wide wall represented by the black segment in
Fig. \ref{fig:dado_bases} (left). It produces a knife-edge effect on
the power emitted by the antennas at a height of $20$m. The
simulated tests presented here account for the shadowing at ground
level.


\subsection{Spectrum cartography}
When designing the basis functions in \eqref{eq:basis-expansion2},
it is known a priori that the transmitted signals are indeed
normalized raised cosine pulses with roll-off factors
$\rho\in\{0,1\}$, and bandwidths $W\in\{10,20,30\}$ MHz. However,
the actual combination of bandwidths and roll-off factors used can
be unknown, which justifies why an overcomplete set of bases becomes
handy. Transmitted signals with bandwidth $W=10$ MHz are searched
over a grid of $16$ evenly spaced center frequencies $f_c$ in $B$.
Likewise, for $W=20$ and $30$ MHz, $15$ and $14$ center frequencies
are considered, respectively. This amounts to $2\times
(16+15+14)=90$ possible combinations for $\rho$, $W$, and $f_c$,
thus $N_b=90$ bases are adopted.

\begin{figure}[h]
\begin{minipage}[b]{.48\linewidth}
  \centering
  \centerline{\includegraphics[width=7cm]{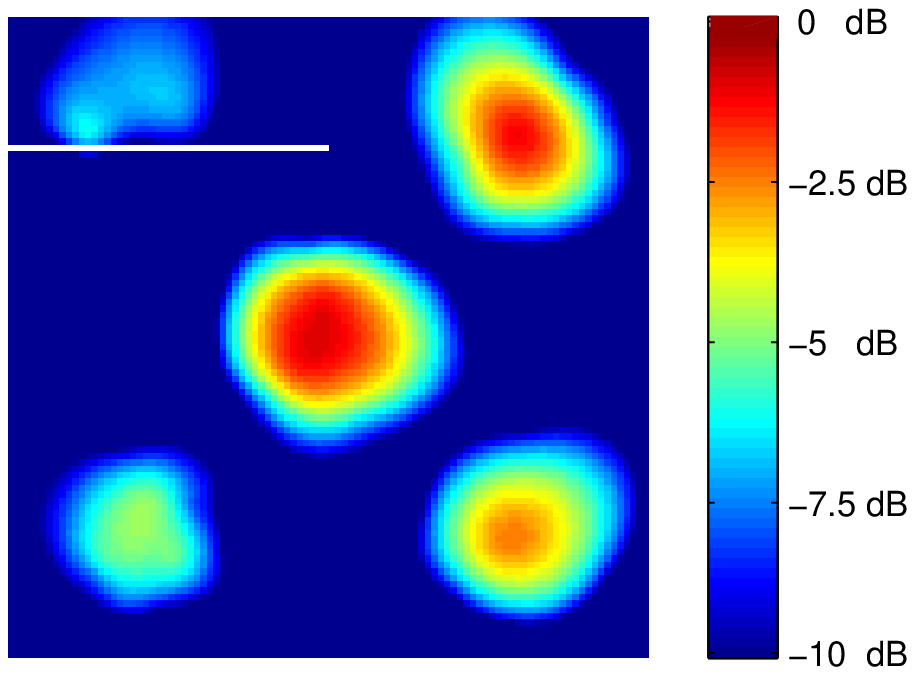}}
 \vspace{-0.8cm}
\medskip
\end{minipage}
\hfill
\begin{minipage}[b]{0.48\linewidth}
  \centering
  \centerline{\includegraphics[width=8cm]{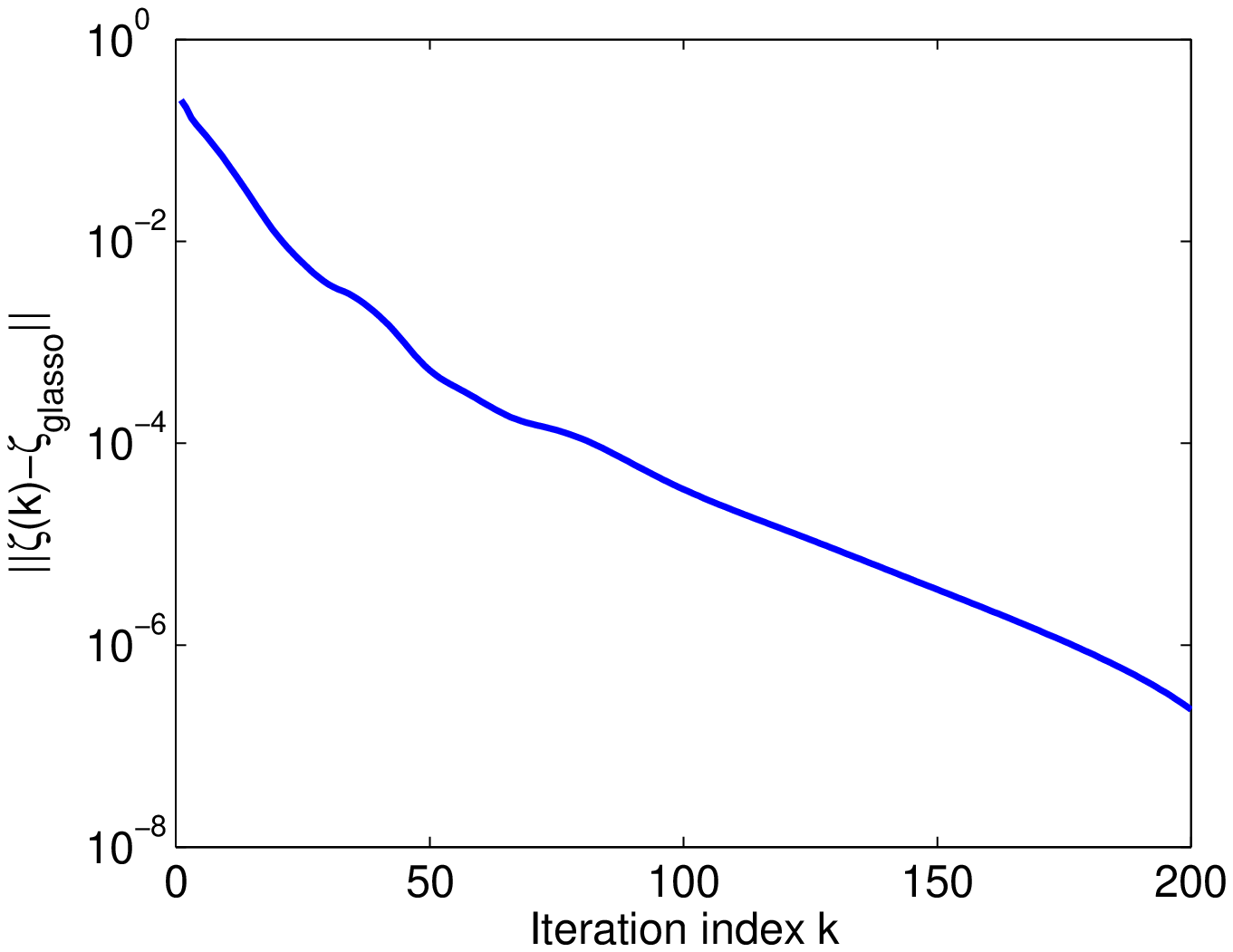}}
 \vspace{-0.8cm}
\medskip
\end{minipage}

\caption{(left) Aggregate map estimate in dB; (right) error
evolution of the GLasso updates} \label{fig:huevos_glasso}
\end{figure}

\begin{figure}[h]
\begin{minipage}[b]{.48\linewidth}
  \centering
  \centerline{\includegraphics[width=8cm]{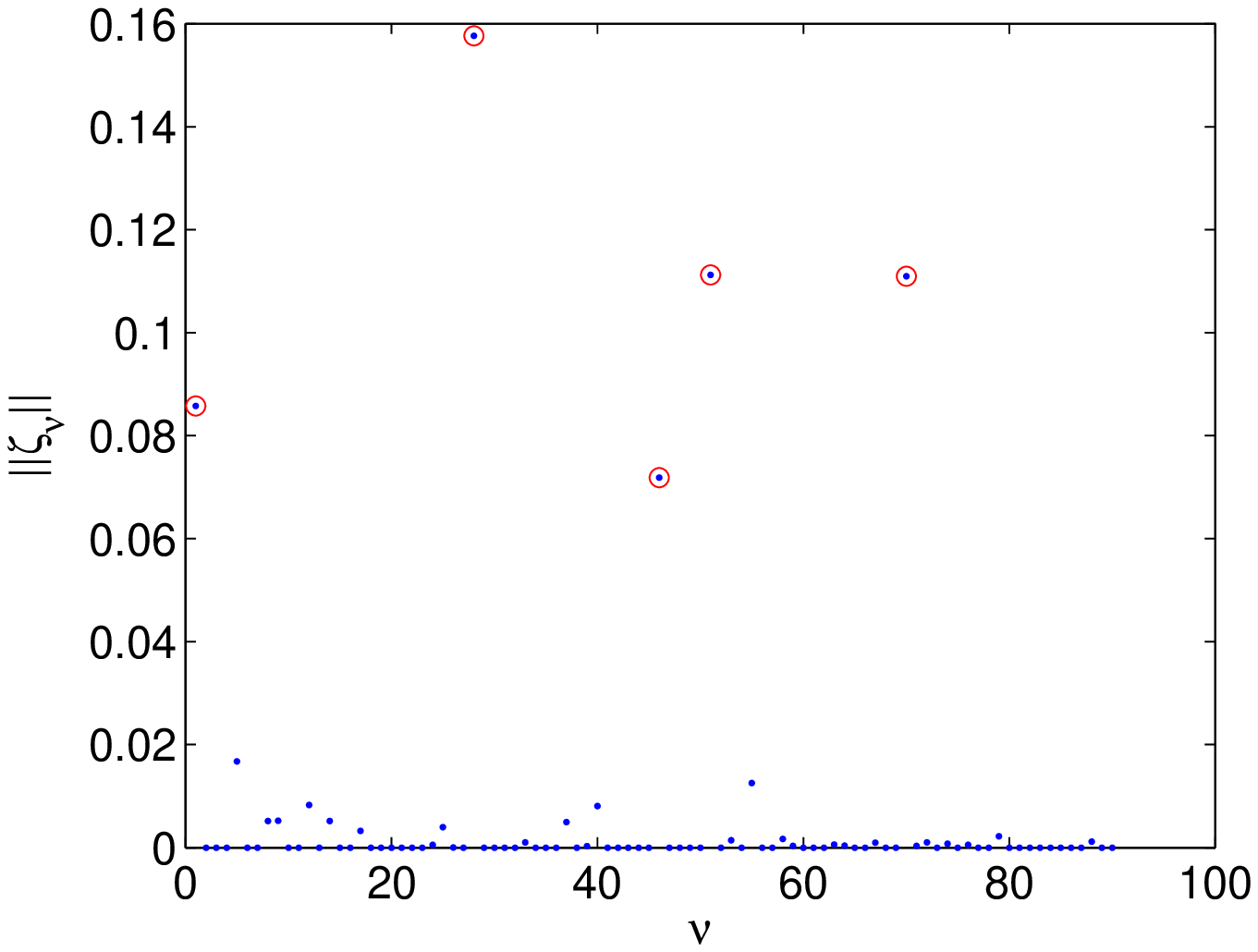}}
 \vspace{-0.8cm}
\end{minipage}
\hfill
\begin{minipage}[b]{0.48\linewidth}
  \centering
  \centerline{\includegraphics[width=\linewidth]{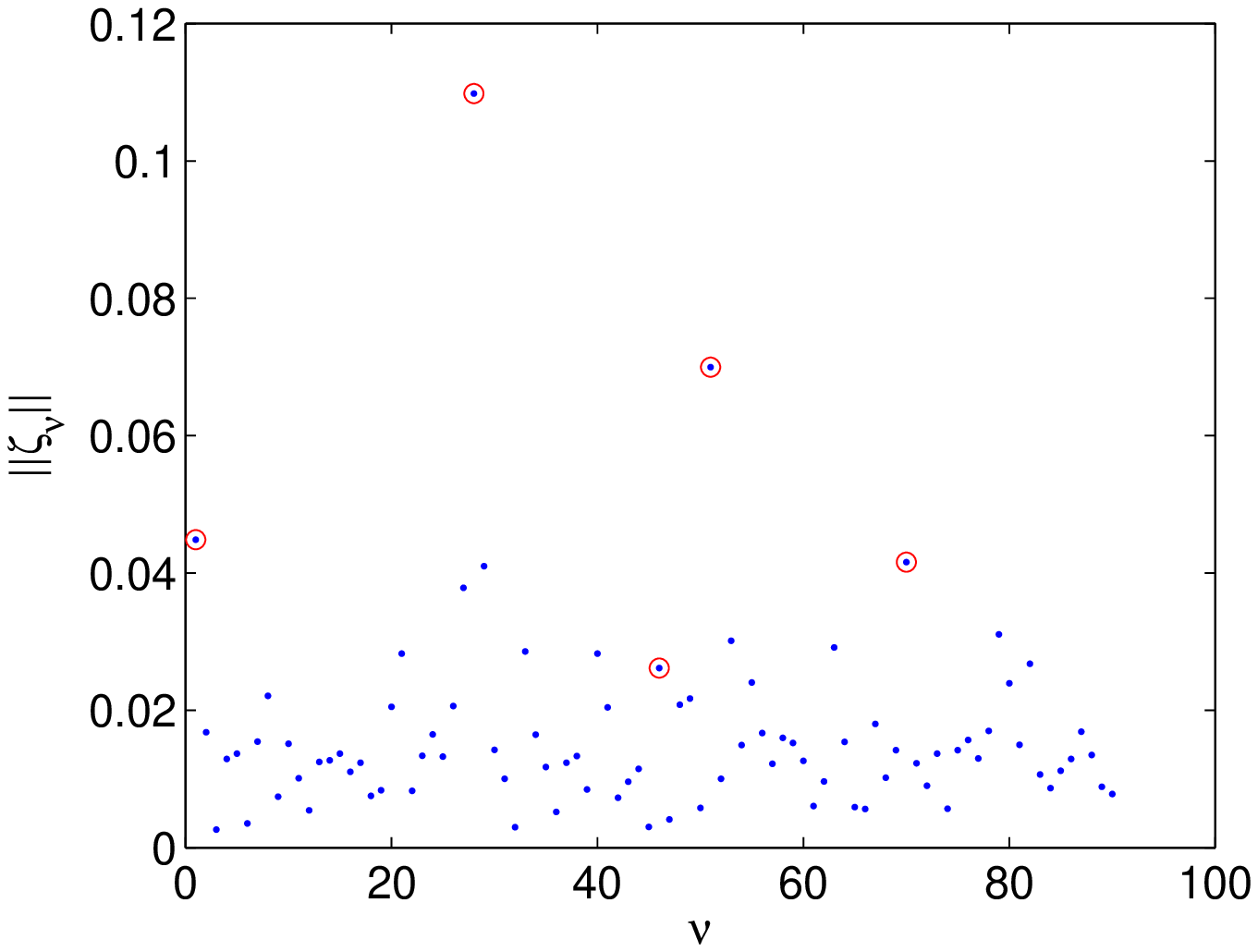}}
 \vspace{-0.8cm}
\end{minipage}

\caption{(left) Frequency bases selected by the group-Lassoed
spline-based estimator; (right) and by ridge regression.}
\label{fig:zetas_path} \vspace{-2cm}

\end{figure}

Each CR computes periodogram samples $\hat\phi_{rn}(\tau)$ at
$N=64$ frequencies with $SNR=-5$ dB, and averages them across
$T=100$ time-slots to form $\varphi_{rn},\ n=1,\ldots,64$ as in
\eqref{eq:intro-ewma}. These network-wide observations at $T=100$
are collected in $\bm \varphi$, and following steps S1-S4 at the
end of Section \ref{sec:BPSC}, the spline-based estimator
\eqref{penalized_LS_GL}, and thus the PSD map $\hat\Phi(\mathbf
x,f)$ is formed. This map is summed across frequencies, and the
result is shown in Fig. \ref{fig:huevos_glasso} (left) which
depicts the positions of transmitting CRs, as well as the
radially-decaying spectra of four of them (those not affected by
the obstacle). It also identifies the effect of the wall by
``flattening'' the spectrum emitted by the fifth source at the
top-left corner. Inspection of the estimate $\hat\Phi(\mathbf
x,f)$ across frequency confirms that group-Lasso succeeds in
selecting the candidate bases. Fig. \ref{fig:zetas_path} (left)
shows points representing $\|\hat{\bm \zeta}_\nu\|_2$,
$\nu=1,\ldots,N_b$, where $\hat{\bm \zeta}_\nu$ is the sub-vector
in the solution of the group-Lasso estimator
\eqref{group_lasso_orig} associated with $g_\nu(\mathbf x)$ and
$b_\nu(f)$. They peak at indexes $\nu=1,\ 28,\ 46,\ 51,$ and $70$
(circled in red), which correspond to the ``ground-truth'' model,
since bases $b_1,\ b_{28},\ b_{46},\ b_{51},$ and $b_{70}$ match
the spectra of the transmitted signals. Even though approximately
75\% of the variables drop out of the model, some spurious
coefficients are retained and their norms are markedly smaller
than those of the ``ground-truth'' bases. This is expected because
based on finite samples there is no guarantee that group-Lasso
will recover the exact support, in general. Nevertheless, the
effectiveness of group-Lasso in revealing the transmitted bases is
apparent when compared to other regularization alternatives. Fig.
\ref{fig:zetas_path} (right) depicts the counterpart of Fig.
\ref{fig:zetas_path} (left) when using a sparsity-agnostic ridge
regression scheme instead of \eqref{group_lasso_orig}. In this
case, no basis selection takes place, and the spurious factors are
magnified up to a level comparable to three of the ``true'' basis
function $b_\nu(f)$. To the best of our knowledge,  no other basis
selection methods in the literature are applicable to the
nonparametric model \eqref{eq:basis-expansion2} considered here.
In particular, COSSO in \cite{cosso} is not applicable since it
does not provide a basis selection method and relies on
orthogonality assumptions.

In summary, this test case demonstrates that the spline-based
estimator can reveal which frequency bands are (un)occupied at each
point in space, thus allowing for spatial reuse of the idle bands.
For instance, transmitter $\textrm{TX}_5$ at the top-right corner is
associated with the basis function $b_{46}(f)$, the only one of the
transmitted five that occupies the $230-260$ MHz sub-band.
Therefore, this sub-band can be reused at locations $\mathbf x$ away
from the transmission range of $\textrm{TX}_5$, which is revealed in
Fig. \ref{fig:huevos_glasso} (left).

The group-Lasso estimator in S1 was obtained via the GLasso
algorithm developed in Section \ref{sec:dglasso} (cf. Algorithm
\ref{GLasso_algorithm_table}). The GLasso output at iteration
$k=1,000$ is compared to previous iterates $\bm \zeta(k)$ in Fig.
\ref{fig:huevos_glasso} (right), which demonstrates the monotone
decay of their difference, and thus corroborates convergence to a
limit point. Then, it is verified numerically that $\bm\zeta(1000)$
satisfies the necessary and sufficient conditions for optimality of
\eqref{group_lasso_dist}, as given in \cite{yuan_group_lasso}. These
two tests together provide numerical confirmation of Proposition
\ref{prop:convergence_DGL} on the convergence of GLasso, and the
optimality of the limit point.

\subsection{Tuning parameters via cross-validation}
Results in Figs. \ref{fig:huevos_glasso} (left) and
\ref{fig:zetas_path} depend on the judicious selection of parameters
$\lambda$ and $\mu$ in \eqref{penalized_LS_GL}. Parameter $\lambda$
affects smoothness, which translates to congruence among PSD
samples, allowing the CRs to recover the radial aspect of the
transmit-power. Parameter $\mu$ controls the sparsity in the
solution, which dictates the number of bases, and thus transmission
schemes that the estimator considers active.

\begin{figure}[h]
\begin{minipage}[b]{.48\linewidth}
  \centering
  \centerline{\includegraphics[width=8cm]{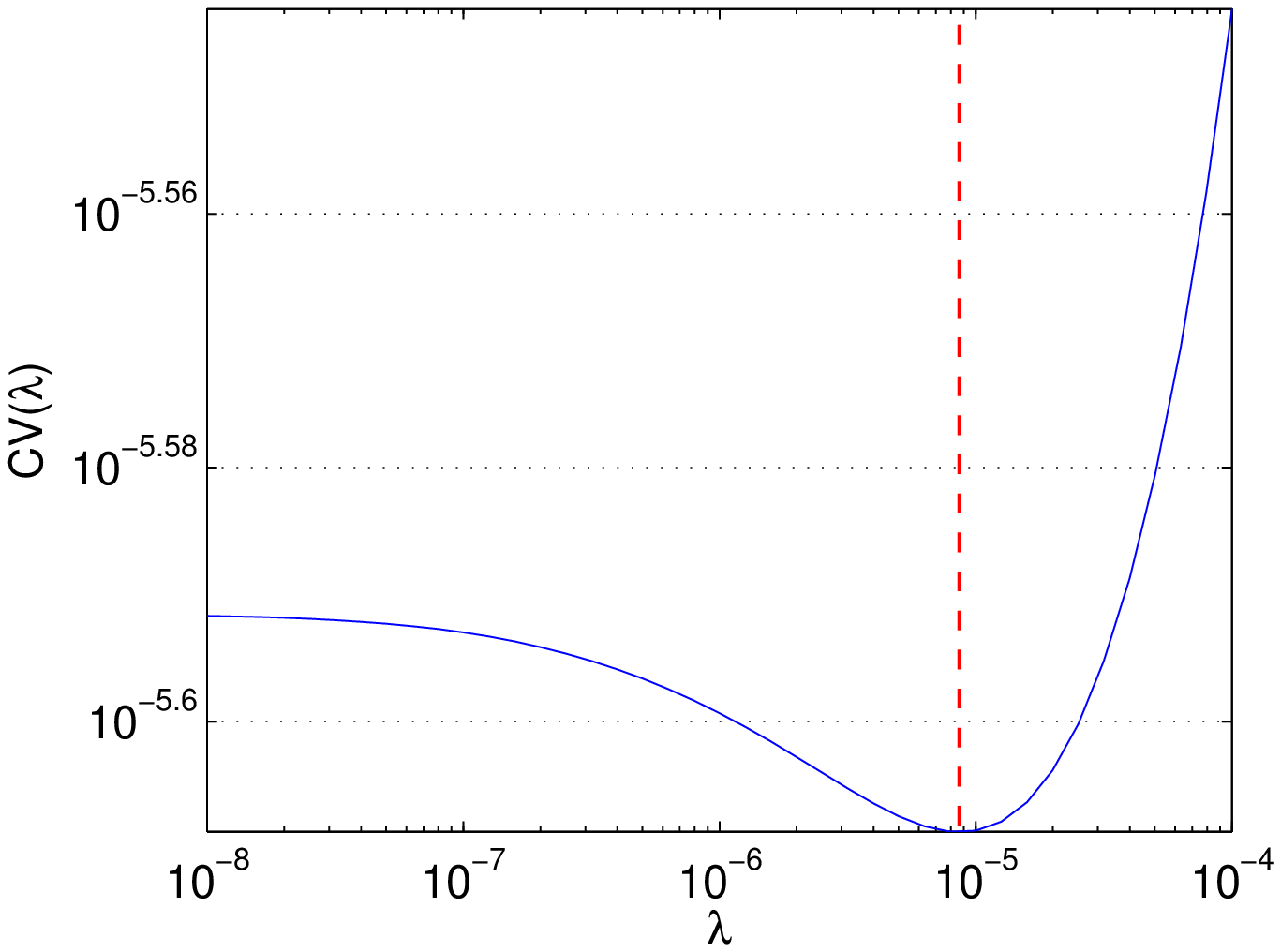}}
 \vspace{-0.8cm}
\end{minipage}
\hfill
\begin{minipage}[b]{0.48\linewidth}
  \centering
  \centerline{\includegraphics[width=8cm]{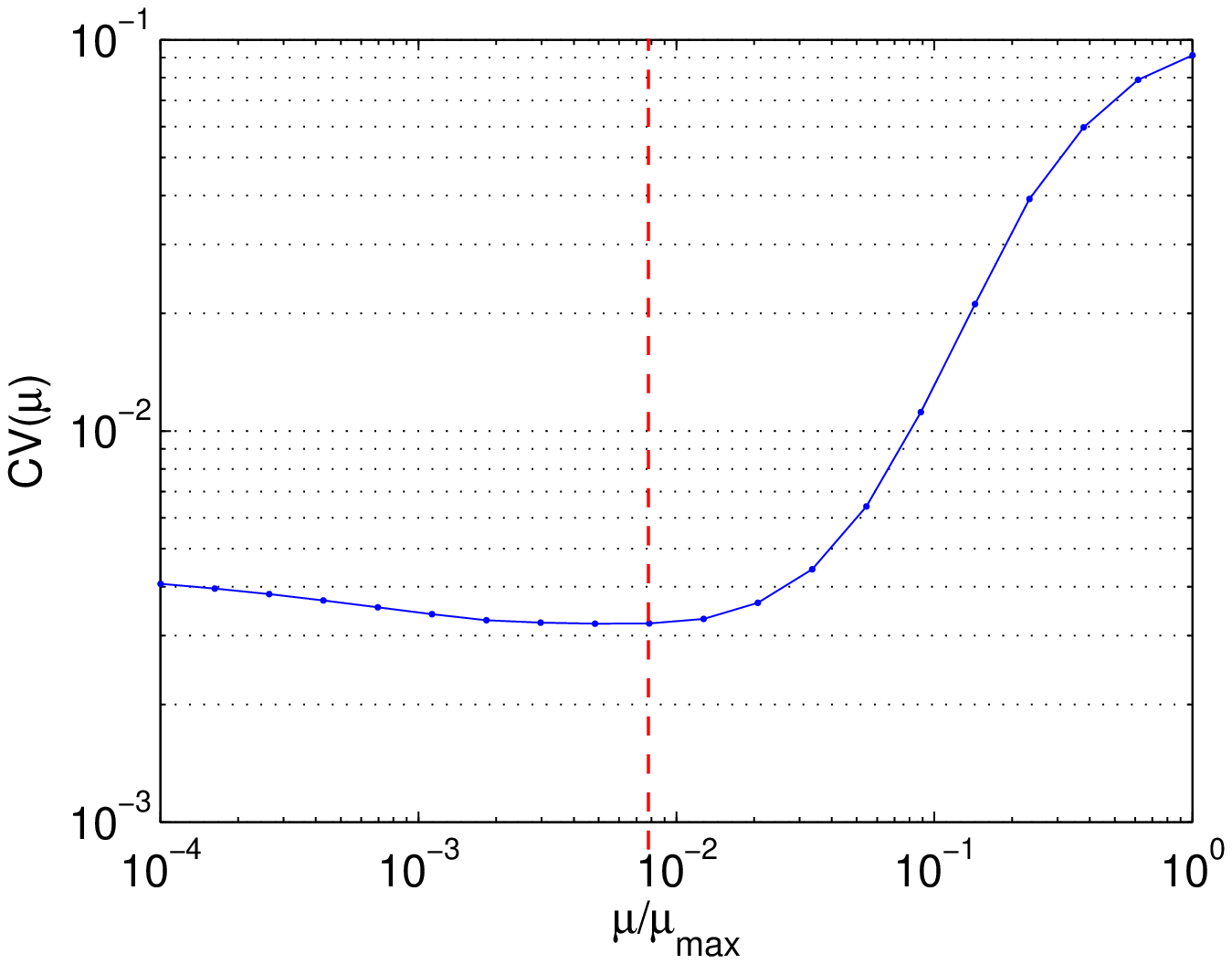}}
 \vspace{-0.8cm}
\end{minipage}

\caption{(left) Minimization of the CV error over $\lambda$;
(right) and over $\mu$.} \label{fig:lambdas_mus} \vspace{5cm}
\end{figure}

To select $\lambda$ and $\mu$ jointly so that both smoothness and
sparsity are properly accounted for, one could consider a
two-dimensional grid of candidate pairs, and minimize the CV error
over this grid. However, this is computationally demanding,
especially because the nondifferentiable cost in
\eqref{group_lasso_orig} renders the shortcuts in Appendix D not
applicable (see also Remark \ref{remark:differences}). A three-step
alternative is followed here. First, estimator
\eqref{penalized_LS_GL} is obtained using an arbitrarily small value
of $\lambda=1\times10^{-6}$, and selecting $\mu=0.1\mu_{\max}$,
where $\mu_{\max}$ is given in subsection \ref{ssec:GL}. In the
second step, only the surviving bases are kept, and the sparsifying
penalty is no longer considered, thus reducing the estimator to that
of Section \ref{sec:batch}. If the reduced matrix $\mathbf B$, built
from the surviving bases, is full rank (otherwise repeat the first
step with a larger value of $\mu$), the procedure in Appendix D is
followed to adjust the value of  $\lambda$ via leave-one-out CV. The
result of this step is illustrated in Fig. \ref{fig:lambdas_mus}
(left), where the minimizer $\lambda_{CV}=7.9433\times 10^{-6}$ of
the OCV cost is selected. The final step consists of reconsidering
the sparsity enforcing penalty in \eqref{penalized_LS_GL}, and
selecting $\mu$ using $5$-fold CV. The minimizer of the CV error
$\mu_{CV}=0.0078\mu_{\max}$ corresponding to this step is depicted
in Fig. \ref{fig:lambdas_mus} (right). Using the $\lambda_{CV}$ and
$\mu_{CV}$ so obtained, the PSD map plotted in Fig.
\ref{fig:huevos_glasso} (left) was constructed. The rationale behind
this approach is that it corresponds to a single step of a
coordinate descent algorithm for minimizing the CV error
$CV(\lambda,\mu)$. Function $CV(\lambda,\mu)$ is typically unimodal,
with much higher sensitivity on  $\mu$ than on $\lambda$, a
geometric feature leading the first coordinate descent update to be
close to the optimum.

\begin{figure}[h]
\vspace{-8cm}
\begin{minipage}[b]{.48\linewidth}
  \centering
  \centerline{\includegraphics[width=8cm]{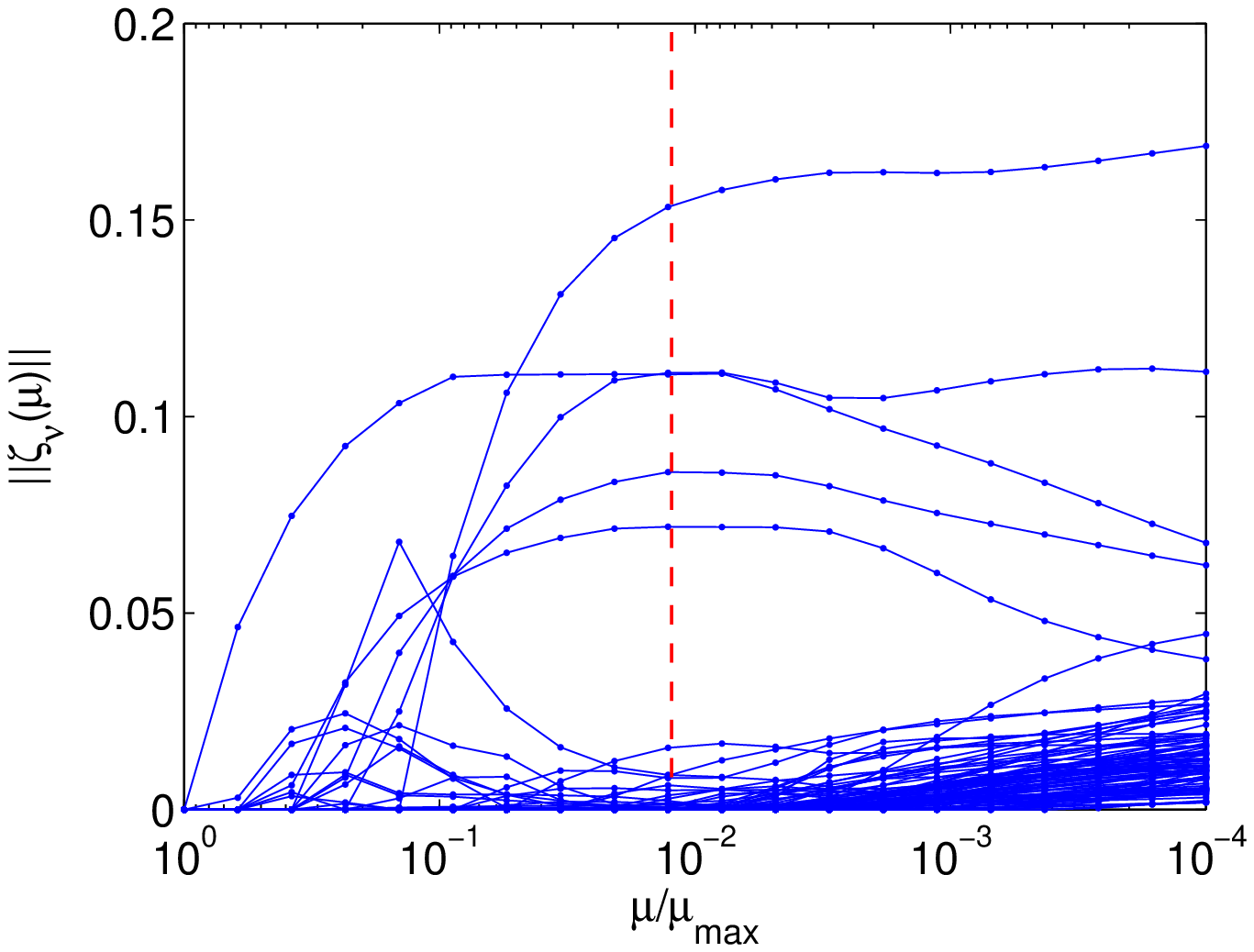}}
 \vspace{-0.8cm}
\end{minipage}
\hfill
\begin{minipage}[b]{0.48\linewidth}
  \centering
  \centerline{\includegraphics[width=\linewidth]{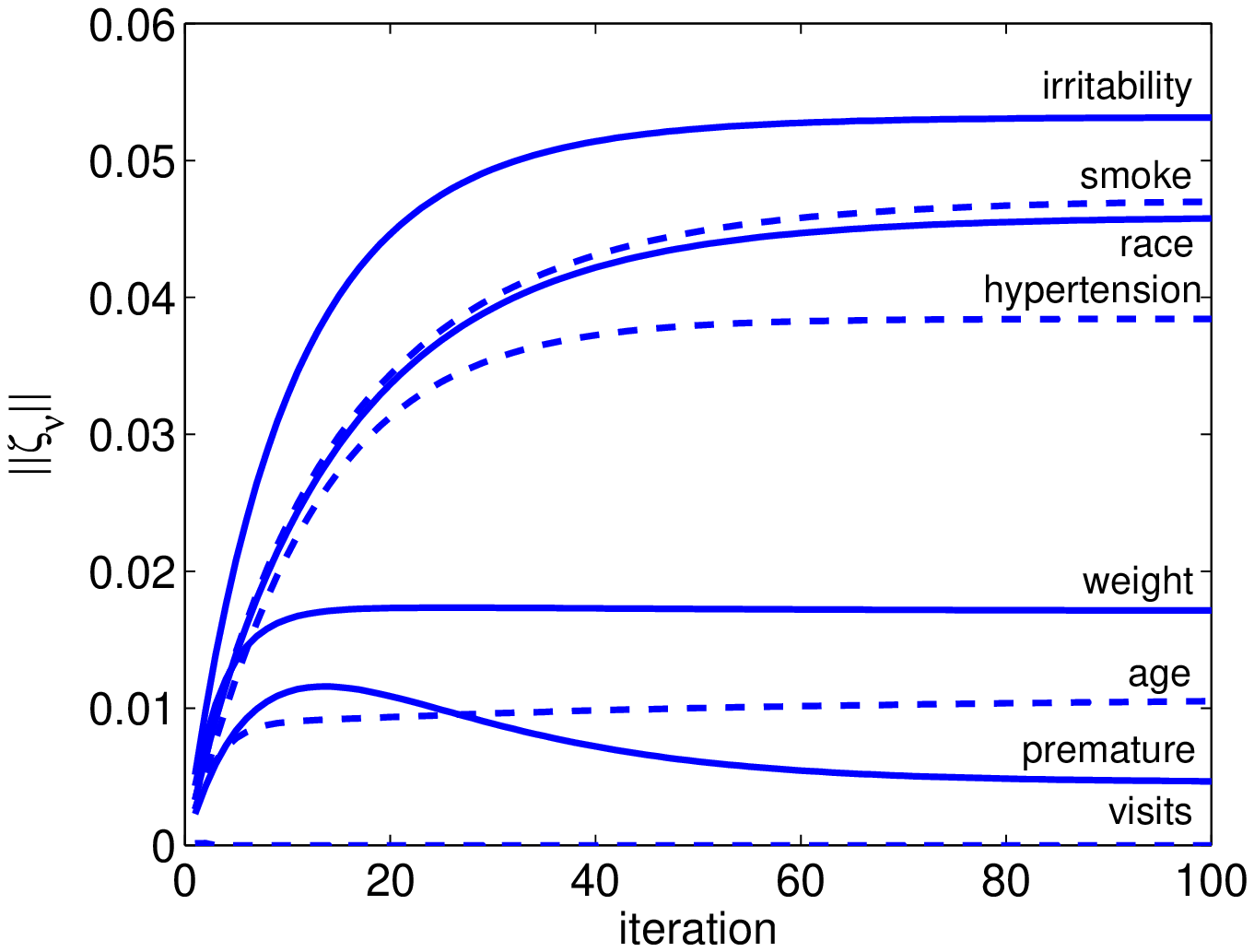}}
 \vspace{-0.8cm}

\end{minipage}

\caption{(left)  Group-Lasso path of solutions $\|\bm\zeta_\nu\|_2$
as $\mu$ varies; (right) factors affecting birthweight, evolution of
GLasso iterates.} \vspace{-2cm} \label{path_birthweight}
\end{figure}

The importance of an appropriate $\mu$ value becomes evident when
inspecting how many bases are retained by the estimator as $\mu$
decreases from $\mu_{\max}$ to $1\times10^{-4}\mu_{\max}$. The
$N_b$ lines in Fig. \ref{path_birthweight} (left) link points
representing $\|\hat{\bm\zeta}_\nu(\mu)\|_2$, as $\mu$ takes on
$20$ evenly spaced values on a logarithmic scale, comprising the
so-termed group-Lasso \textit{path of solutions}. When
$\mu=\mu_{\max}$ is selected, by definition the estimator forces
all $\hat{\bm\zeta}_\nu$ to zero, thus discarding all bases. As
$\mu$ tends to zero all bases become relevant and eventually enter
the model, which confirms the premise that LS estimators suffer
from overfitting when the underlying model is overcomplete. The
cross-validated value $\mu_{CV}$ is indicated with a dashed
vertical line that crosses the path of solutions at the values of
$\|\hat{\bm\zeta}_\nu\|_2$. At this point, five sub-vectors
corresponding to the factors $\nu=1,\ 28,\ 46,\ 51,$ and $70$ are
considerably far away from zero hence showing strong effects, in
par with the results depicted in Fig. \ref{fig:zetas_path} (left).
Certainly interesting would be to corroborate the effectiveness of
the proposed PSD map estimator on real data comprising spatially
distributed RF measurements. Upon availability of such dataset,
this direction will be pursued and reported elsewhere.

\subsection{Example with real data}

The goal of this section is to demonstrate that the GLasso
algorithm in Section \ref{sec:dglasso} can be useful for
applications other than the spline-based BEM for spectrum
cartography dealt with in Sections \ref{sec:batch} and
\ref{sec:BPSC}. This demonstration will rely on the birthweight
dataset from \cite{homster}, considered also by the seminal
group-Lasso work of \cite{yuan_group_lasso}. The objective is to
predict the human birthweight from $p=8$ factors including the
mother's \verb"age", \verb"weight", \verb"race", \verb"smoke"
habits, number of previous \verb"premature" labors, history of
\verb"hypertension", uterine \verb"irritability", and number of
physician \verb"visits" during the first trimester of pregnancy.
Third-order polynomials were considered to model nonlinear effects
of the \verb"age" and \verb"weight" on the response, augmenting
the model size to $p=12$ by grouping the polynomial coefficients
in two subsets of three variables.

GLasso is run under this setup, over the set of $N=189$ samples,
with $\mu$ selected via 7-fold CV. Fig. \ref{path_birthweight}
(right) depicts the evolution of the factors' strength measured by
$\|\zeta_\nu\|_2^2$, which -- as expected -- converge to the same
prediction model as in \cite{yuan_group_lasso}. Additionally, GLasso
is capable of determining that the eighth factor (\verb"visits") is
not significant even from the first iterations, allowing for early
model selection.

\section{Concluding Summary}\label{sec:conclusion}
A basis expansion approach was introduced in this paper to estimate
a multi-dimensional field, whose dependence on a subset of its
variables is modeled through preselected (and generally
\emph{overlapping}) basis functions weighted by unknown
coefficient-functions of the remaining variables. The unknown
coefficient functions can be estimated from the field's noisy
samples, by solving a variational LS problem which admits infinite
solutions. Without extra constraints, the estimated field
interpolates perfectly the data samples, at the price of severely
overfitting the true field elsewhere. The first contribution was to
regularize this variational LS cost by a smoothing term, which can
afford a unique finite-parameter spline-based solution. The latter
is expressed in terms of radial kernels and polynomials whose
parameters were estimated in closed form. A recursive PSD tracker
was also developed for slowly time-varying spectra.

The second main contribution pertains to a robust variant of the
function estimator, when an overcomplete set of bases is adopted to
effectively accommodate model uncertainties. The novel estimator
here minimizes the variational LS cost regularized by a term that
performs \emph{basis selection}, and thus yields a parsimonious
description of the field by retaining those few members of the basis
that ``better'' explain the data. This attribute is achieved because
the added penalty induces a group (G)Lasso estimator on the
parameters of the kernels and polynomials.  Even though the number
of unknowns increases with overcomplete bases, most coefficients are
zero, meaning that the complexity remains at an affordable level
using the sparsity-promoting GLasso. Notwithstanding, (group-) Lasso
here is introduced to effect (group-) sparsity in the space of
smooth functions.

The third contribution is a provably convergent GLasso estimator
developed based on AD-MoM iterations. It entails parallel
\emph{closed-form} updates, which involve simple vector
soft-thresholding operations per factor. Its fully-distributed
counterpart was also developed for use by a network of wireless
sensors, or, multiple processors to balance the load of a
computational cluster. It is worth stressing that both GLassso and
DGLasso are standalone tools for sparse linear regression,
applicable to a gamut of problems that go beyond the field
estimation context of this paper.

The fourth contribution is in the context of wireless CR network
sensing (our overarching practical motivation), where the
overcomplete estimated field enables cartographing the
space-frequency distribution of power generated by RF sources whose
transmit-PSDs are shaped by, e.g., raised-cosine pulses with
possibly different roll-off factors, center frequencies, and
bandwidths. Using periodogram samples collected by spatially
distributed CRs, the sparsity-aware spline-based estimator yields an
atlas of \textit{PSD maps} (one map per frequency). As corroborated
by simulations, the atlas enables localizing the sources and
discerning their transmission parameters, even in the presence of
frequency-selective Rayleigh fading and pronounced shadowing effects
due to e.g., an obstructing wall. Simulated tests also illustrated
the convergence of Glasso, and confirmed that the sparsity-promoting
regularization is effective in selecting those basis functions that
strongly influence the field, when the tuning parameters are
cross-validated properly.

Given the existing connections between splines and classical
estimators for both random and deterministic field models, the
spline-based methods developed in this paper provide a unifying
framework suitable for both paradigms. The model and the
resultant (parsimonious) estimates can thus be used in more
general statistical inference and localization problems,
whenever the data admit a basis expansion over a proper subset
of its dimensions. Furthermore, results in this paper extend to
kernels other than radial basis functions, whenever the
smoothing penalty is replaced by a norm induced from an RKHS.
Also of interest is to quantify the number of data required to
attain a prescribed approximation error, in light of the
existing connections between spline-based reconstruction and
Shannon's sampling theory~\cite{unser_tutorial}.

\section*{Acnowledgments} The authors would like to thank Prof. Hui
Zou (School of Statistics, University of Minnesota) for his
feedback which improved the exposition of ideas in this paper.

{\Large\appendix}

\noindent\normalsize \emph{\textbf{A. Proof of Proposition
\ref{prop:thin-splines}}}: 
%
Rewrite \eqref{penalized_LS} as
\begin{equation}\label{penalized_LS3}
\min_{\{g_\nu\in\mathcal{S}\}_{\nu=2}^{N_b}}\left[\min_{g_1\in\mathcal{S}}
\sum_{r=1}^{N_r}\sum_{n=1}^{N} \left(\varphi_{rn}^{(-1)}-
g_1(\mathbf{x}_r)b_1(f_n)\right)^2+ \lambda \int_{\mathbb{R}^2}
||\nabla^2 g_1(\mathbf{x})||_F^2 d\mathbf{x}\right]+ \lambda
\sum_{\nu=2}^{N_b}\int_{\mathbb{R}^2} ||\nabla^2
g_\nu(\mathbf{x})||_F^2 d\mathbf{x}
\end{equation}
with $\varphi_{rn}^{(-1)}:=\varphi_{rn}-\sum_{\nu=2}^{N_b}
g_\nu(\mathbf{x}_r)b_\nu(f_n)$. Focusing on the inner minimization
w.r.t. $g_1$, fix the set of functions $\{g_\nu\}_{\nu=2}^{N_b}$,
and note that only the first two terms are relevant (those within
the square brackets). It follows from \cite[Theorem 4 bis]{duchon}
that $\hat g_1$ takes the form in \eqref{eq:claim}, with
coefficients $\bm\beta_1, \bm \alpha_{11}$ and $\alpha_{10}$ that
depend on $G^{(-1)}:=\{g_\nu(x_r),\ r=1,\ldots,N_r,\
\nu=2,\ldots,N_b\}$ through $\varphi_{rn}^{(-1)}$. 
The next step is to minimize \eqref{penalized_LS3} w.r.t. $g_2$
but with $\{g_\nu\}_{\nu=3}^{N_b}$ fixed, which amounts to
\begin{equation}\label{penalized_LS4}
\min_{g_2\in\mathcal{S}} \sum_{r=1}^{N_r}\sum_{n=1}^{N}
\left(\varphi_{rn}^{(-2)}- \hat
g_{1}(\mathbf{x}_r)b_1(f_n)-g_2(\mathbf{x}_r)b_2(f_n)\right)^2+\lambda
\int_{\mathbb{R}^2} ||\nabla^2 \hat g_{1}(\mathbf{x})||_F^2
d\mathbf{x}+ \lambda \int_{\mathbb{R}^2} ||\nabla^2
g_2(\mathbf{x})||_F^2 d\mathbf{x}
\end{equation}
where $\varphi_{rn}^{(-2)}:=\varphi_{rn}-\sum_{\nu=3}^{N_b}
g_\nu(\mathbf{x}_r)b_\nu(f_n)$. In the first two summands of the
cost in \eqref{penalized_LS4}, $\hat g_1$ depends on $g_2$ via
$G^{(-1)}$. Because $G^{(-1)}$ only involves evaluating $g_2$ on
$\mathcal X$, \cite[Theorem 4 bis]{duchon} can be applied again,
and the optimal solution $\hat g_2$ takes the from
\eqref{eq:claim}. The same argument carries over to subsequent
minimization steps for $\nu=3,\ldots,N_b$, establishing that all
$\{\hat g_\nu(\mathbf{x})\}$ are thin-plate splines as in
\eqref{eq:claim}.


\noindent\normalsize \emph{\textbf{B. Proof of
\eqref{system_eq1}-\eqref{system_eq3}}}:
Upon substituting \eqref{eq:claim} into \eqref{penalized_LS}, it
will shown next that the optimal coefficients
$\{\hat{\bm\alpha},\hat{\bm\beta}\}$ specifying
$\{\hat{g}_{\nu}(\mathbf{x})\}_{\nu=1}^{N_b}$ are obtained as
solutions to the following constrained, regularized LS problem
\begin{align}\label{constr_LS}
&\min_{\bm
\alpha,\bm\beta}\frac{1}{{N_rN}}\left\|\bm\varphi-(\mathbf{B}\otimes\mathbf{K})\bm\beta-
(\mathbf{B}\otimes \mathbf{T})\bm
\alpha\right\|_{2}^{2}+\lambda\bm\beta^{\prime}(\mathbf{I}_{N_b}
\otimes \mathbf{K})\bm\beta\nonumber\\
&{\quad}\mbox{s.
t.}{\quad}(\mathbf{I}_{N_b}\otimes\mathbf{T}^{\prime})\bm\beta=\mathbf{0}_{3N_b}.
\end{align}

Observe first that the constraints $\bm \beta_\nu\in\mathcal{B}$
in Proposition 1 can be expressed as $
\mathbf{T}^{\prime}\bm\beta_\nu=\mathbf{0}_3$ for each
$\nu=1,\ldots,N_b$, or jointly as
$(\mathbf{I}_{N_b}\otimes\mathbf{T}^{\prime})\bm\beta=\mathbf{0}_{3N_b}$.
For the optimization objective in \eqref{constr_LS}, note from
\eqref{eq:claim} that
$\hat{g}_\nu(\mathbf{x}_r)=\mathbf{k}_r^\prime\bm\beta_\nu+\mathbf{t}_r^\prime\bm\alpha_\nu$,
where $\mathbf{k}_r^\prime$ and $\mathbf{t}_r^\prime$ are the
$r$th rows of $\mathbf K$ and $\mathbf T$, respectively. The first
term in the cost of \eqref{penalized_LS} can be expressed  (up to
a factor $({N_rN})^{-1}$) as
\begin{eqnarray}
\sum_{n=1}^{N}\sum_{r=1}^{N_r}
\left(\varphi_{rn}-\sum_{\nu=1}^{N_b}b_\nu(f_n)[\mathbf{k}_r^\prime\bm\beta_\nu
+\mathbf{t}_r^\prime\bm\alpha_\nu] \right)^2 &=&
\sum_{n=1}^{N}\sum_{r=1}^{N_r}
\left(\varphi_{rn}-(\mathbf{b}_n\otimes\mathbf{k}_r)^\prime\bm\beta
-(\mathbf{b}_n\otimes\mathbf{t}_r)^\prime\bm\alpha \right)^2\nonumber\\
&=& \sum_{n=1}^{N}
\left\|\bm\varphi_{n}-(\mathbf{b}_n^\prime\otimes\mathbf{K})\bm\beta
-(\mathbf{b}_n^\prime\otimes\mathbf{T})\bm\alpha \right\|_2^2\nonumber\\
&=&\left\|\bm\varphi-(\mathbf{B}\otimes\mathbf{K})\bm\beta
-(\mathbf{B}\otimes\mathbf{T})\bm\alpha \right\|_2^2\nonumber.
\end{eqnarray}
Consider next the  penalty term in the cost of
\eqref{penalized_LS}. Substituting into \eqref{eq:claim}, it
follows that $\int_{\mathbb{R}^2} ||\nabla^2
\hat{g}_\nu(\mathbf{x})||_F^2
d\mathbf{x}=\bm\beta_\nu^\prime\mathbf{K}\bm\beta_\nu$ ~\cite[p.
33]{wahba}. It thus holds that
\begin{equation*}
\lambda \sum_{\nu=1}^{N_b}\int_{\mathbb{R}^2} ||\nabla^2
\hat{g}_\nu(\mathbf{x})||_F^2 d\mathbf{x}=\lambda
\sum_{\nu=1}^{N_b}\bm\beta_\nu^\prime
\mathbf{K}\bm\beta_\nu=\lambda\bm\beta^{\prime}(\mathbf{I}_{N_b}
\otimes \mathbf{K})\bm\beta
\end{equation*}
from which \eqref{constr_LS} follows readily.

Now that the equivalence between \eqref{penalized_LS} and
\eqref{constr_LS} has been established, the latter must be solved
for $\bm \alpha$ and $\bm \beta$. Even though $\mathbf{K}$ (hence
$\mathbf{I}_{N_b} \otimes \mathbf{K}$) is not positive definite,
it is still possible to show that
$\bm\beta^{\prime}(\mathbf{I}_{N_b} \otimes \mathbf{K})\bm\beta>0$
for any $\bm \beta$ such that
$(\mathbf{I}_{N_b}\otimes\mathbf{T}^{\prime})\bm\beta=\mathbf{0}_{3N_b}$~\cite[p.
85]{duchon}, implying that \eqref{constr_LS} is convex. Proceeding
along the lines of~\cite[p. 33]{wahba}, note first that the
constraint
$(\mathbf{I}_{N_b}\otimes\mathbf{T}^{\prime})\bm\beta=\mathbf{0}_{3N_b}$
implies the existence of a vector
$\bm\gamma\in\mathbb{R}^{N_{b}(N_r-3)}$ satisfying
\eqref{system_eq3}. After this change of variables,
\eqref{constr_LS} is transformed into an unconstrained quadratic
program, which can be solved in closed form for
$\{{\bm\alpha},{\bm\gamma}\}$. Hence, setting both gradients
w.r.t. ${\bm\alpha}$ and ${\bm\gamma}\}$ to zero yields
\eqref{system_eq1} and \eqref{system_eq2}.

\noindent\normalsize \emph{\textbf{C. Proof of Proposition
\ref{prop:thin-splines-GL}}}:
After substituting \eqref{parametric_expansion_GL} into
\eqref{penalized_LS_GL}, one finds the optimal
$\{{\bm\alpha},{\bm\beta}\}$ specifying
$\{\hat{g}_{\nu}(\mathbf{x})\}_{\nu=1}^{N_b}$ in
\eqref{parametric_expansion_GL},  as solutions to the following
constrained, regularized LS problem
\begin{align}\label{constr_LS_GL}
&\min_{\bm
\alpha,\bm\beta}\frac{1}{{N_rN}}\left\|\bm\varphi-(\mathbf{B}\otimes\mathbf{K})\bm\beta-
(\mathbf{B}\otimes \mathbf{T})\bm
\alpha\right\|_{2}^{2}+\lambda\bm\beta^{\prime}(\mathbf{I}_{N_b}
\otimes \mathbf{K})\bm\beta+\mu\sum_{\nu=1}^{N_b}\|\mathbf{K}\bm\beta_\nu+\mathbf{T}\bm\alpha_\nu\|_2\nonumber\\
&{\quad}\mbox{s.
t.}{\quad}(\mathbf{I}_{N_b}\otimes\mathbf{T}^{\prime})\bm\beta=\mathbf{0}_{3N_b}.
\end{align}

With reference to \eqref{constr_LS_GL}, consider grouping and
reordering the variables $\{\bm\alpha,\bm\beta\}$ in the vector
$\mathbf{u}:=[\mathbf{u}_1^{\prime},\ldots,\mathbf{u}_{N_b}^{\prime}]^{\prime}$,
where
$\mathbf{u}_\nu:=[\bm\beta_\nu^{\prime}\:\:\bm\alpha_\nu^{\prime}]^{\prime}$.
As argued in Section \ref{ssec:solution}, the constraints
$\mathbf{T}^{\prime}\bm\beta_\nu=\mathbf{0}$ can be eliminated
through the change of variables
$\mathbf{u}_\nu=\textrm{bdiag}(\mathbf{Q}_2,\mathbf{I}_3)\mathbf{v}_\nu$
for $\nu=1,\ldots,N_b$; or compactly as
$\mathbf{u}=(\mathbf{I}_{N_b}\otimes\textrm{bdiag}(\mathbf{Q}_2,\mathbf{I}_3))\mathbf{v}$.
The next step is to express the three summands in the cost of
\eqref{constr_LS_GL} in terms of the new vector optimization
variable $\mathbf{v}$. Noting that $\mathbf{k}_r^\prime\bm\beta_\nu
+\mathbf{t}_r^\prime\bm\alpha_\nu=[\mathbf{k}_r^\prime\:\:
\mathbf{t}_r^\prime]\mathbf{u}_\nu$, and mimicking the steps in
Appendix A, the first summand is
\begin{align}\label{first_term_LS}
\frac{1}{{N_rN}}\left\|\bm\varphi-(\mathbf{B}\otimes\mathbf{K})\bm\beta
-(\mathbf{B}\otimes\mathbf{T})\bm\alpha \right\|_2^2
=&\:\frac{1}{{N_rN}}\left\|\bm\varphi-(\mathbf{B}\otimes[\mathbf{K}\:\:\mathbf{T}])\mathbf{u}\right\|_2^2\nonumber\\
=&\:\frac{1}{{N_rN}}\left\|\bm\varphi-(\mathbf{B}\otimes[\mathbf{KQ}_2\:\:\mathbf{T}])\mathbf{v}\right\|_2^2.
\end{align}
The second summand due to the thin-plate penalty can be
expressed as
\begin{align}\label{second_term_LS}
\lambda\sum_{\nu=1}^{N_b}\bm\beta_\nu^\prime\mathbf{K}\bm\beta_\nu=&\:
\lambda\sum_{\nu=1}^{N_b}\mathbf{u}_\nu^\prime\textrm{bdiag}(\mathbf{K},\mathbf{0})\mathbf{u}_\nu=
\lambda\sum_{\nu=1}^{N_b}\mathbf{v}_\nu^\prime\textrm{bdiag}(\mathbf{Q}_2^\prime\mathbf{K}
\mathbf{Q}_2,\mathbf{0})\mathbf{v}_\nu\nonumber\\
=&\:
\lambda\mathbf{v}^\prime(\mathbf{I}_{N_b}\otimes\textrm{bdiag}(\mathbf{Q}_2^\prime\mathbf{K}
\mathbf{Q}_2,\mathbf{0}))\mathbf{v}
\end{align}
while the last term is
$\mu\sum_{\nu=1}^{N_b}\|\mathbf{K}\bm\beta_\nu+\mathbf{T}\bm\alpha_\nu\|_2=
\mu\sum_{\nu=1}^{N_b}\|[\mathbf{K}\:\:\mathbf{T}]\mathbf{u}_\nu\|_2=
\mu\sum_{\nu=1}^{N_b}\|[\mathbf{KQ}_2\:\:\mathbf{T}]\mathbf{v}_\nu\|_2.$
Combining \eqref{first_term_LS} with \eqref{second_term_LS} by
completing the squares, problem \eqref{constr_LS_GL} is
equivalent to
\begin{equation}\label{almost_group_lasso}
\min_{\mathbf{v}}\frac{1}{N_r N}\left\|\left[\begin{array}{c}\bm\varphi\\
\mathbf{0}\end{array}\right]-
\left[\begin{array}{c}\mathbf{B}\otimes[\mathbf{KQ}_2\:\:\mathbf{T}]\\
\mathbf{I}_{N_b}\otimes\textrm{bdiag}(({N_rN}\lambda\mathbf{Q}_2^{\prime}\mathbf{K}\mathbf{Q}_2
)^{1/2},\mathbf{0})\end{array}\right]\mathbf{v}\right\|_{2}^{2}
+\mu\sum_{\nu=1}^{N_b}\|[\mathbf{KQ}_2\:\:\mathbf{T}]\mathbf{v}_\nu\|_2
\end{equation}
and becomes \eqref{group_lasso_orig}  under  the identities
\eqref{eq:identity_yX}, and after the change of variables
$\bm\zeta:=[\bm\zeta_1^{\prime},\ldots,\bm\zeta_{N_b}^{\prime}]^{\prime}=
(\mathbf{I}_{N_b}\otimes[\mathbf{KQ}_2\:\:\mathbf{T}])\mathbf{v}$.
By definition of $\mathbf{u}$, $\mathbf{v}$, and $\bm\zeta$, the
original variables can be recovered through the transformation in
\eqref{change_of_variables_GL}.

\noindent\normalsize \emph{\textbf{D. Selection of the smoothing
parameter in \eqref{penalized_LS}}}:  The method to be developed
builds on the so-termed leave-one-out CV, which proceeds as
follows; see e.g.,~\cite[Ch. 4]{wahba}. Consider removing a single
data point $\varphi_{rn}$ from the collection of ${N_rN}$
measurements available to the sensing radios. For a given
$\lambda$, let $\hat\Phi^{(-rn)}_\lambda(\mathbf{x},f)$ denote the
\textit{leave-one-out} estimated PSD map, obtained by solving
\eqref{penalized_LS} following steps S1-S3 in Section
\ref{ssec:solution}, using the ${N_rN}-1$ remaining data points.
 The aforementioned estimation procedure is repeated
$N_r N$ times by leaving out each of the data points
$\varphi_{rn},$ $r=1,\ldots,N_r$ and $n=1,\ldots,N$, one at a
time. The leave-one-out or ordinary CV (OCV)~\cite[p.
242]{elements_of_statistics},~\cite[p. 47]{wahba}, for the problem
at hand is given by
\begin{equation}\label{ocv_function}
\textrm{OCV}(\lambda)=\frac{1}{{N_rN}}\sum_{r=1}^{N_r}
\sum_{n=1}^N\left(\varphi_{rn}-\hat\Phi^{(-rn)}_\lambda(\mathbf{x}_r,f_n)\right)^2
\end{equation}
while the optimum $\lambda$ is selected as the minimizer of
$\textrm{OCV}(\lambda)$, over a grid of values
$\lambda\in[0,\lambda_{\max}]$. Function \eqref{ocv_function}
constitutes an average of the squared prediction errors over all
data points; hence, its minimization offers a natural criterion.
The method is quite computationally demanding though, since the
system of linear equations \eqref{system_eq1}-\eqref{system_eq3}
has to be solved $N_r N$ times for each value of $\lambda$ on the
grid. Fortunately, this computational burden can be significantly
reduced for the spline-based PSD map estimator considered here.

Recall the vector $\bm\varphi$ collecting all data points measured
at locations $\mathcal{X}$ and frequencies $\mathcal{F}$. Define
next a similar vector $\hat{\bm\varphi}$ containing the respective
predicted values at the given locations and frequencies, which is
obtained after solving \eqref{penalized_LS} with all the data in
$\bm\varphi$ and a given value of $\lambda$. The following lemma
establishes that the PSD map estimator is a \textit{linear
smoother}, which means that the predicted values are linearly
related to the measurements, i.e.,
$\hat{\bm\varphi}=\mathbf{S}_\lambda\bm\varphi$ for a
$\lambda$-dependent matrix $\mathbf{S}_\lambda$ to be determined.
Common examples of linear smoothers are ridge regressors and
smoothing splines; further details are in~\cite[p.
153]{elements_of_statistics}. For linear smoothers, by virtue of
the leave-one-out lemma~\cite[p. 50]{wahba} it is possible to
rewrite \eqref{ocv_function} as
\begin{equation}\label{ocv_function_cheap}
\textrm{OCV}(\lambda)= \frac{1}{{N_rN}}\sum_{r=1}^{N_r}
\sum_{n=1}^N\left(\frac{\varphi_{rn}-\hat\Phi_\lambda(\mathbf{x}_r,f_n)}{1-[\mathbf{S}_\lambda]_{ii}}\right)^2
\end{equation}
where $\hat\Phi_\lambda(\mathbf{x},f)$ stands for the estimated
PSD map when all data in $\bm\varphi$ are utilized in
\eqref{penalized_LS}. The beauty of the leave-one-out lemma stems
from the fact that  given $\lambda$ and the main diagonal of
matrix $\mathbf{S}_\lambda$, the right-hand side of
\eqref{ocv_function_cheap} indicates that fitting a single model
(rather than $N_r N$ of them) suffices to evaluate
$\textrm{OCV}(\lambda)$. The promised lemma stated next specifies
the value of $\mathbf{S}_\lambda$ necessary to evaluate
\eqref{ocv_function_cheap}.

\begin{lemma}\label{lemma:smoother}
The PSD map estimator in \eqref{penalized_LS} is a linear
smoother, with smoothing matrix given by
\begin{align}\label{smoothing_matrix}
\mathbf{S}_\lambda=&\:(\mathbf{B}\otimes
\{\mathbf{KQ}_2-\mathbf{TR}^{-1}\mathbf{Q}_1^{\prime}
\mathbf{KQ}_2\})[(\mathbf{B}^{\prime}\mathbf{B} \otimes
\mathbf{Q}_2^\prime\mathbf{K}\mathbf{Q}_2)+{N_rN}\lambda
\mathbf{I}]^{-1}(\mathbf{B}^{\prime}\otimes
\mathbf{Q}_2^{\prime})\nonumber\\
&\:+(\mathbf{B}\bm\Gamma^{-1}\bm\Omega_1^{-1}\otimes\mathbf{TR}^{-1}\mathbf{Q}_1^\prime).
\end{align}
\end{lemma}
\begin{IEEEproof}
Reproduce the structure of $\bm\varphi$ in Section
\ref{ssec:solution} to form the supervector
$\hat{\bm\varphi}:=[\hat{\bm\varphi}_1^{\prime},\ldots,\hat{\bm\varphi}_N^{\prime}]^{\prime}\in\mathbb{R}^{{N_rN}}$,
by stacking each vector
$\hat{\bm\varphi}_n:=[\hat{\Phi}_\lambda(\mathbf{x}_1,f_n),\ldots,\hat{\Phi}_\lambda(\mathbf{x}_{N_r},f_n)]^{\prime}$
corresponding to the spatial PSD predictions at frequency
$f_n\in\mathcal{F}$. From \eqref{eq:claim}, it follows that
$\hat{\Phi}_\lambda(\mathbf{x}_r,f_n)=(\mathbf{b}_n\otimes\mathbf{k}_r)^\prime\hat{\bm\beta}
-(\mathbf{b}_n\otimes\mathbf{t}_r)^\prime\hat{\bm\alpha}$, where
$\mathbf{b}_n^\prime$, $\mathbf{k}_r^\prime$ and
$\mathbf{t}_r^\prime$ are the $n$th and $r$th rows of $\mathbf B$,
$\mathbf K$ and $\mathbf T$, respectively. By stacking the PSD map
estimates, it follows that
$\hat{\bm\varphi}_n=(\mathbf{b}_n^\prime\otimes\mathbf{K})\hat{\bm\beta}
-(\mathbf{b}_n^\prime\otimes\mathbf{T})\hat{\bm\alpha},$ which
readily yields
\begin{equation}\label{predicted_values}
\hat{\bm\varphi}=(\mathbf{B}\otimes\mathbf{K})\hat{\bm\beta}
-(\mathbf{B}\otimes\mathbf{T})\hat{\bm\alpha}.
\end{equation}
Because the estimates $\{\hat{\bm\alpha},\hat{\bm\beta}\}$ are
linearly related to the measurements $\bm\varphi$ [cf.
\eqref{system_eq1}-\eqref{system_eq3}], so is $\hat{\bm\varphi}$
as per \eqref{predicted_values}, establishing that the PSD map
estimator in \eqref{penalized_LS} is indeed a linear smoother.
Next, solve explicitly for $\{\hat{\bm\alpha},\hat{\bm\beta}\}$ in
\eqref{system_eq1}-\eqref{system_eq3} and substitute the results
in \eqref{predicted_values}, to unveil the structure of the
smoothing matrix $\mathbf{S}_\lambda$ such that
$\hat{\bm\varphi}=\mathbf{S}_\lambda\bm\varphi$. Simple algebraic
manipulations lead to the expression \eqref{smoothing_matrix}.
\end{IEEEproof}

The effectiveness of the leave-one-out CV approach is corroborated
via simulations in Section \ref{sec:sims}.

\noindent\normalsize \emph{\textbf{E. Proof of
\eqref{pupdate_1}-\eqref{gammaupdate_1}}:}
\label{appx:equations_admom} Recall the augmented Lagrangian
function in \eqref{Aug_Lagr_1}, and let
$\bm\zeta:=\{\bm\zeta_r\}_{r\in\calR}$ for notational brevity. When
used to solve \eqref{constrmin_1}, the three steps in the AD-MoM are
given by:
\begin{description}
\item [{\bf [S1]}]  \textbf{Local estimate updates:}
    \begin{equation}\label{S1_1}\bm\zeta(k+1)=
    \mbox{arg}\:\min_{\bm\zeta}\ccalL_{c}\left[\bm\zeta,
    \bm\gamma(k),\calbbv(k)\right].
    \end{equation}

\item [{\bf [S2]}] \textbf{Auxiliary variable updates:}
    \begin{equation}\label{S2_1}\bm\gamma(k+1)=
    \mbox{arg}\min_{\bm{\gamma}}\ccalL_{c}\left[\bm\zeta(k+1),
    \bm\gamma,\calbbv(k)\right].
    \end{equation}
\item [{\bf [S3]}] \textbf{Multiplier
    updates:}\begin{align}
    \bbv_{r}(k+1)&=\bbv_{r}(k)+c[\bm\zeta_{r}(k+1)-\bm\gamma_r(k+1)]\label{S3_a_1}\\
\breve\bbv_{r}^{r^\prime}(k+1)&=\breve\bbv_{r}^{r^\prime}(k)+c[\bm\zeta_r(k+1)-{\bm\gamma}_{r}^{r^\prime}(k+1)]\label{S3b1}\\
    \bar{\bbv}_{r}^{r^\prime}(k+1)&=\bar{\bbv}_{r}^{r^\prime}(k)+c[\bm\zeta_{r^\prime}(k+1)-{\bm\gamma}_{r}^{r^\prime}(k+1)].\label{S3c1}
\end{align}
\end{description}

Focusing first on [S2], observe that \eqref{Aug_Lagr_1} is
separable across the collection of variables $\{\bm\gamma_j\}$ and
$\{{\bm\gamma}_r^{r^\prime}\}$ that comprise $\bm\gamma$. The
minimization w.r.t. the latter group reduces to
\begin{align}
\nonumber{\bm\gamma}_r^{r^\prime}(k+1)&=\arg\min_{{\bm\gamma}_r^{r^\prime}} c\|{\bm\gamma}_r^{r^\prime}\|^2 - c\Big(\bm \zeta_r(k+1)+\bm\zeta_{r^\prime}(k+1)\Big){\bm\gamma}_r^{r^\prime}-\Big(\bar{ \bbv}_r^{r^\prime}(k)+\breve{ \bbv}_r^{r^\prime}(k)\Big){\bm\gamma}_r^{r^\prime}\\
\nonumber&=\frac{1}{2}\Big(\bm \zeta_r(k+1)+\bm\zeta_{r^\prime}(k+1)\Big)+\frac{1}{2c}\Big(\bar{\bbv}_r^{r^\prime}(k)+\breve{\bbv}_r^{r^\prime}(k)\Big)\\
&=\frac{1}{2}\Big(\bm
\zeta_r(k+1)+\bm\zeta_{r^\prime}(k+1)\Big)\label{gammarr}.
\end{align}

The result in \eqref{gammarr} assumes that
$\bar{\bbv}_r^{r^\prime}(k)+\breve{\bbv}_r^{r^\prime}(k)=\mathbf{0},\
\forall k$.
 A simple inductive argument over \eqref{S3b1}, \eqref{S3c1} and \eqref{gammarr} shows that this is indeed true if the multipliers
 are initialized such that $\bar{\bbv}_r^{r^\prime}(0)+\breve{\bbv}_r^{r^\prime}(0)=\mathbf{0}$.

The remaining minimization in \eqref{S2_1} with
respect to $\{\bm\gamma_r\}$ decouples into $N_r$ quadratic
sub-problems [cf. \eqref{Aug_Lagr_1}], that is
\begin{equation*}
\bm\gamma_r(k+1)=\mbox{arg}\:\min_{\bm\gamma_r}
\frac{1}{2}\left\|\mathbf{y}_r-\mathbf{X}_{r}\bm\gamma_{r}\right\|_2^{2}
-\bbv_r^\prime(k)\bm\gamma_r+\frac{c}{2}
\|\bm\zeta_{r}(k+1)-\bm\gamma_r\|_2^{2}
\end{equation*}
which admit the closed-form solutions in \eqref{gammaupdate_1}.

In order to obtain the update \eqref{pupdate_1}  for the prices
$\bbp_r$,
 consider their definition together with
\eqref{S3b1}, \eqref{S3c1} and \eqref{gammarr} to obtain
\begin{align*}
\bbp_r(k+1)&=\sum_{r^\prime\in \mathcal N_r}\Big( \breve\bbv_{r}^{r^\prime}(k+1)+\bar{\bbv}_{r^\prime}^{r}(k+1)\Big)\\
&=\sum_{r^\prime\in \mathcal N_r}\Big( \breve\bbv_{r}^{r^\prime}(k)+\bar{\bbv}_{r^\prime}^{r}(k)\Big)+\sum_{r^\prime\in \mathcal N_r}c
 \Big(2\bm\zeta_{r}(k+1)-\bm\gamma_r^{r^\prime}(k)-\bm\gamma^r_{r^\prime}(k)\Big)\\
&=\bbp_r(k)+c\sum_{r^\prime\in \mathcal
N_r}\left(\bm\zeta_r(k+1)-\bm\zeta_{r^\prime}(k+1)\right)
\end{align*}
which coincides with \eqref{pupdate_1}.

Towards obtaining the updates for the local variables in
$\bm\zeta$, the optimization \eqref{S1_1} in [S1] can be also
split into $N_r$ sub-problems, namely
\begin{align}\label{betamin_1}
\nonumber\bm\zeta_r(k+1)=\mbox{arg}\:\min_{\bm\zeta_r} &{\:}
\left\{\phantom{\sum_{r^\prime\in\calN_r}}\hspace{-0.8cm}\frac{\mu}{N_r}\sum_{\nu=1}^{N_b}\|\bm\zeta_{r\nu}\|_2
+\bbv_r^\prime(k)\bm\zeta_r +\frac{c}{2}
\|\bm\zeta_{r}-\bm\gamma_{r}(k)\|_2^{2}
+\sum_{r^\prime\in\calN_r}\left[\breve\bbv_{r}^{r^\prime}(k)+\bar{\bbv}^{r}_{r^\prime}(k)\right]^\prime
\bm\zeta_{r}\right.\\
\nonumber&\left.\hspace{0.2cm}+\frac{c}{2}\sum_{r^\prime\in\calN_r}
\left[\|\bm\zeta_{r}-{\bm\gamma}_{r}^{r^\prime}(k)\|_2^{2}+
\|\bm\zeta_{r}-{\bm\gamma}^{r}_{r^\prime}(k)\|_2^2\right]\right\}\\
\nonumber=\mbox{arg}\:\min_{\bm\zeta_r} &{\:}\left\{
\phantom{\sum_{r^\prime\in\calN_r}}\hspace{-0.8cm}\frac{\mu}{N_r}\sum_{\nu=1}^{N_b}\|\bm\zeta_{r\nu}\|_2
-\left(c\sum_{r^\prime\in \mathcal N_r}\Big(
\bm\zeta_{r}(k)+\bm\zeta_{r^\prime}(k)\Big)+c\bm\gamma_r(k)-\mathbf{p}_r(k)-\bbv_r(k) \right)^\prime\bm\zeta_r\right.\\
\nonumber&+ \left. \frac{c}{2}(1+2|\mathcal
N_r|)\|\bm\zeta_r\|_2^2\phantom{\sum_{r^\prime\in\calN_r}}\hspace{-.7cm}\right\}.
\end{align}
Upon dividing by $c(1+2|\mathcal N_r|)$ each subproblem becomes
identical to problem \eqref{glasso_tonto}, and thus by
Proposition \ref{prop:glasso_closed_form} takes the form of
\eqref{betaupdate_1}.

\bibliographystyle{IEEEtranS}
\bibliography{IEEEabrv,biblio}


%
%

%

%

\end{document}